\documentclass[twocolumn]{aastex63}
\usepackage{amsmath}
\shorttitle{Revisiting the High-Redshift Lensed Quasars}
\shortauthors{Yue et al.}
\graphicspath{{./}{figures/}}

\begin{document}

\title{Revisiting the Lensed Fraction of High-Redshift Quasars}

\correspondingauthor{Minghao Yue}
\email{yuemh@email.arizona.edu}

\author[0000-0002-5367-8021]{Minghao Yue}
\affiliation{Steward Observatory, University of Arizona, 933 North Cherry Avenue, Tucson, AZ 85721, USA}

\author[0000-0003-3310-0131]{Xiaohui Fan}
\affiliation{Steward Observatory, University of Arizona, 933 North Cherry Avenue, Tucson, AZ 85721, USA}

\author[0000-0001-5287-4242]{Jinyi Yang}
\altaffiliation{Strittmatter Fellow}
\affil{Steward Observatory, University of Arizona, 933 North Cherry Avenue, Tucson, AZ 85721, USA}

\author[0000-0002-7633-431X]{Feige Wang}
\altaffiliation{NHFP Hubble Fellow}
\affiliation{Steward Observatory, University of Arizona, 933 North Cherry Avenue, Tucson, AZ 85721, USA}



\begin{abstract}
The observed lensed fraction of high-redshift quasars $(\sim0.2\%)$
is significantly lower than previous theoretical predictions $(\gtrsim4\%)$.
We revisit the lensed fraction of high-redshift quasars predicted by theoretical models, where we
adopt recent measurements of galaxy velocity dispersion functions (VDFs) 
and explore a wide range of quasar luminosity function (QLF) parameters. 
We use both analytical methods and mock catalogs which give consistent results.
For ordinary QLF parameters and the depth of current high-redshift quasar surveys $(m_z\lesssim22)$, 
our model suggests a multiply-imaged fraction of $F_\text{multi}\sim 0.4\%-0.8\%$.
The predicted lensed fraction is $\sim1\%-6\%$ for the brightest $z_s\sim6$ quasars $(m_z\lesssim19)$,
depending on the QLF.
The systematic uncertainties of the predicted lensed fraction in previous models
can be as large as $2-4$ times and
 are dominated by the VDF. Applying VDFs from recent measurements decreases the predicted lensed fraction and  relieves the tension between observations and theoretical models. 
Given the depth of current imaging surveys, there are $\sim15$ lensed quasars at $z_s>5.5$
 detectable over the sky. Upcoming sky surveys like the LSST survey and the {\em Euclid} survey 
 will find several tens of lensed quasars at this redshift range.

\end{abstract}


\section{Introduction}

In the past two decades, 
extensive searches have been carried out for high-redshift quasars
\citep[e.g.,][]{jiang16,matsuoka18,wang19,wang21,yang19,yang20},
which led to a sample of more than 1200 quasars at $z>4.5$ (more than 500 at $z>5$).
In \citet{fan19}, we reported the discovery of the first gravitationally lensed quasar at $z>5$
(J0439+1634 at $z=6.52$).
High-redshift lensed quasars such as J0439+1634 are valuable, as they
enable in-depth investigations of distant quasars with enhanced sensitivity and resolution
\citep[e.g.,][]{yang19b, yue21}.

Since the commissioning of the Sloan Digital Sky Survey \citep[SDSS,][]{sdss},
there have been efforts to search for lensed quasars at $z\gtrsim4$ \citep[e.g.,][]{richards2004, richards06}.
However, there are only three lensed quasars at $z>4.5$ known so far, including a couple at $z=4.8$ \citep{mcgreer10, more16} and J0439+1634 at $z=6.52$.
The fraction of lensed ones among high-redshift quasars is $\sim0.2\%$ (without correcting for survey incompleteness).
Meanwhile, a number of theoretical studies have predicted a high lensed fraction $(\gtrsim4\%)$ 
for high-redshift quasars due to strong magnification bias
\citep[e.g.,][]{wyithe02,comerford02,pacucci19}.
\citet{fan19} and \citet{pacucci19} suggested that survey incompleteness is a plausible reason for the discrepancy.

We notice, however, that previous models were based on various assumptions on the
quasar population and the deflector population,
which might introduce systematic errors that have not been fully explored.
In particular, many previous studies were based on early measurements of quasar luminosity functions (QLFs)
and deflector velocity dispersion functions (VDFs). 
Since recent observations have provided more accurate measurements of the QLFs and VDFs,
there is a need to re-investigate the models of high-redshift lensed quasars to fully understand the
discrepancy between theoretical predictions and observations.

In this paper, we revisit the models of high-redshift lensed quasar population
and investigate possible systematic uncertainties.
We adopt recent measurements of VDFs and QLFs, and use both analytical methods and mock catalogs to 
model the properties of high-redshift lensed quasars.
Our analysis addresses the following questions:
(1) what are the uncertainties of the theoretical models,
(2) how many lensed quasars can we detect in current and future sky surveys, and
(3) how strong is the tension between the theoretical predictions and the observations?
Answering these questions is critical to 
the design of a complete search for high-redshift lensed quasars
with the upcoming sky surveys like the Vera C. Rubin Observatory Legacy Survey of Space and Time \citep[LSST, ][]{lsst}.

This paper is organized as follows.
We describe the analytical method in Section \ref{sec:analytical} and the mock catalog method in Section \ref{sec:mock}.
We present the predicted statistics of high-redshift lensed quasars in Section \ref{sec:stat},
discuss the implications in Section \ref{sec:discussion},
and summarize our conclusions in Section \ref{sec:conclusion}.
We use a flat $\Lambda$CDM cosmology with $H_0=70\text{ km s}^{-1}\text{Mpc}^{-1}$ and $\Omega_M=0.3$
throughout this work.
All magnitudes are $z-$band AB magnitudes unless further specified.

\section{Analytical Method} \label{sec:analytical}

In this Section, we describe the analytical model of the high-redshift lensed quasar population, 
including the lensing model, the deflector galaxies, and the source quasars.
In this paper, we use ``strongly-lensed" and ``multiply-imaged"
interchangeably.

\subsection{Lensing Models}


We use singular isothermal spheres (SIS) to describe the mass profile of deflector galaxies. 
SIS and singular isothermal ellipsoids \citep[SIE, e.g.,][]{sie} are widely used in models of strong-lensing systems 
\citep[e.g.,][]{wyithe02, om10}.
SIS is parameterized by its Einstein radius, $\theta_E = 4\pi\left(\frac{\sigma}{c}\right) ^2  \frac{D_{ds}}{D_{s}}$,
where $\sigma$ is the velocity dispersion of the deflector, $D_{ds}$ and $D_{s}$ are the angular diameter distances 
from the source to the deflector and the observer, respectively.
Following \citet{wyithe11},
the lensing optical depth of a population of SIS deflectors, 
i.e., the probability for a source at redshift $z_s$ to be multiply-imaged, is
\begin{equation}\label{eq:tm}
    \tau_m=\int_0^{z_s} dz_d \int d\sigma \phi(\sigma, z_d) \frac{d^2V_c}{d\Omega dz_d} \pi \theta_E(\sigma, z_d, z_s)^2 D_d^2
\end{equation}
where $z_d$ is the deflector redshift, $\phi(\sigma, z_d)$ is the deflector VDF, 
$\frac{d^2V_c}{d\Omega dz_d}=(1+z_d)^3c\frac{dt}{dz_d}$ is the differential comoving volume,
and $D_d$ is the angular diameter distance at $z_d$.



The lensing optical depth $\tau_m$ describes the fraction of 
lensed ones among {\em all} background sources.
In flux-limited surveys, however, 
some lensed sources can get detected which 
 are intrinsically fainter than the survey limit.
Consequently,
we expect that the observed lensed fraction in flux-imited surveys
should be higher than $\tau_m$.
The magnification bias $B$ quantifies this effect, 
\begin{equation} \label{eq:B}
    B=\frac{\int_{\mu_\text{min}}^{+\infty} d\mu ~ p(\mu) N(>L_\text{lim}/\mu)}{N(>L_\text{lim})}
\end{equation}
where $p(\mu)$ is the probabilistic distribution of the magnification of lensed sources, 
$L_\text{lim}$ is the survey flux limit,
and $N(>L_\text{lim})$ is the number of background sources brighter than $L_\text{lim}$.
According to Equation \ref{eq:B}, if there are 
$N(>L_\text{lim})$ strongly-lensed objects that are intrinsically brighter than $L_\text{lim}$, the number of strongly-lensed objects that have magnified flux brighter than $L_\text{lim}$ should be
$B\times N(>L_\text{lim})$.

In our analytical model, we use $\mu_\text{min}=2$  which
is the minimum magnification for multiply-imaged systems generated by SIS.
If the survey has a high spatial resolution which can resolve the lensing structure,
we shall use $p(\mu)=2/(\mu-1)^3$ which describes the magnification of the brighter lensed image.
Similarly, for surveys with poor resolution where the lensed images are blended,
we use $p(\mu)=8/\mu^3$ which corresponds to the total magnification.
\footnote{We derive $p(\mu)$ from Equation 8.34(c) -- 8.35(b) 
from \citet{gravlens}. Also see \citet{wyithe11}.}
Although SIS deflectors cannot produce quad lenses,
the fraction of quad lenses among all lenses is small \citep[$\sim10\%$, e.g.,][]{om10}.

The fraction of lensed sources among all sources that have {\em apparent} luminosity brighter than $L_\text{lim}$ is 
\citep[e.g.,][]{mason15}:
\begin{equation} \label{eq:fmulti}
    F_\text{multi} = \frac{B \tau_m}{B \tau_m + B'(1-\tau_m)}
\end{equation}
where $B'$ is the magnification bias of sources that are not multiply imaged. 
The major source of $B'$ is weak lensing.
As shown in \citet{mason15}, for sources at $z_s>6$,
the magnification caused by weak lensing $(\mu_\text{weak})$ are distributed within a narrow range around $\mu_\text{weak}\approx1$,
with almost all sources have $0.7<\mu_\text{weak}<1.3$.
According to Equation \ref{eq:B}, we use $B'\approx1$ in our model.
This approximation is also adopted in \citet{wyithe11}.

\subsection{Deflector Population}

\begin{figure*}
    \centering
    \includegraphics[width=0.42\linewidth]{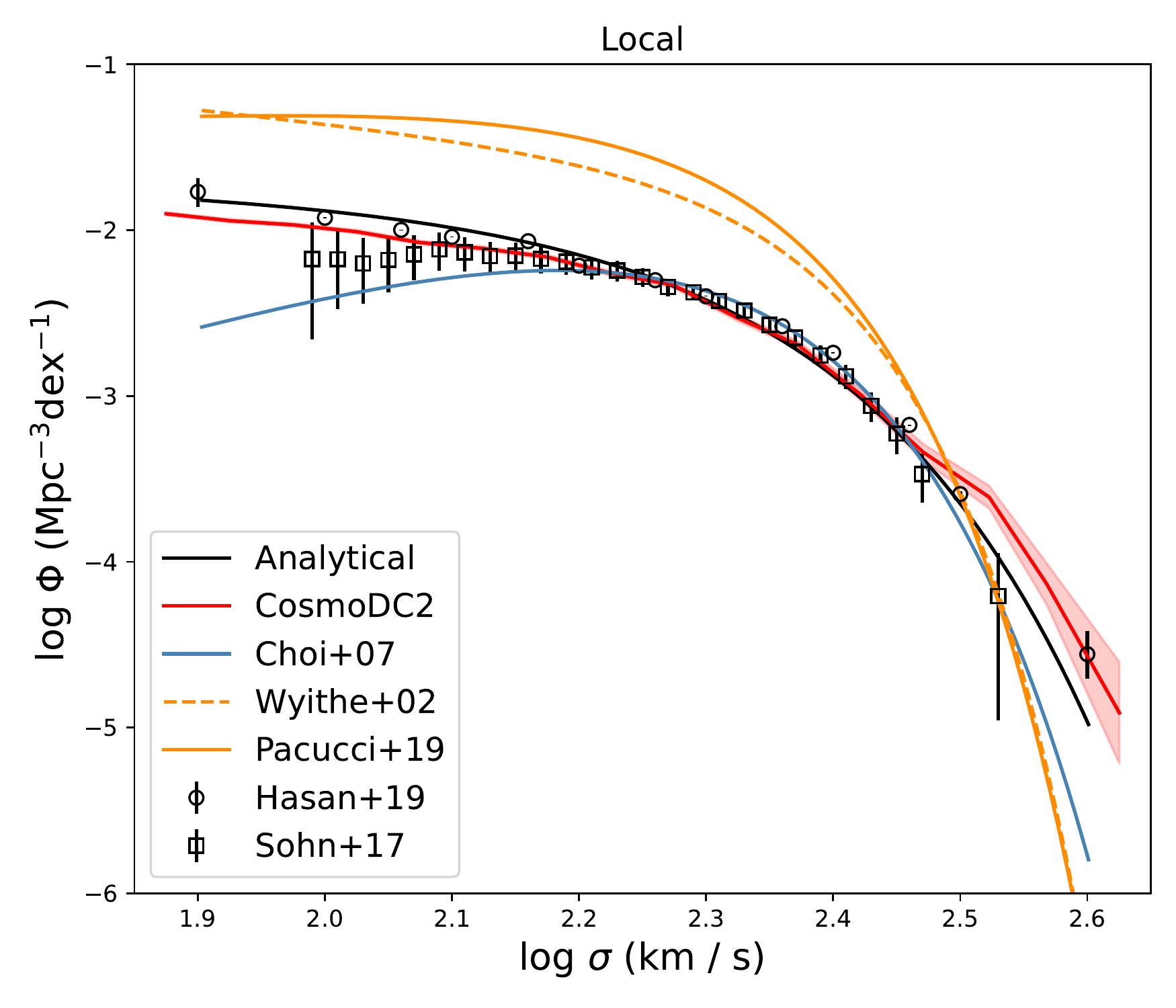}
    \includegraphics[width=0.48\linewidth]{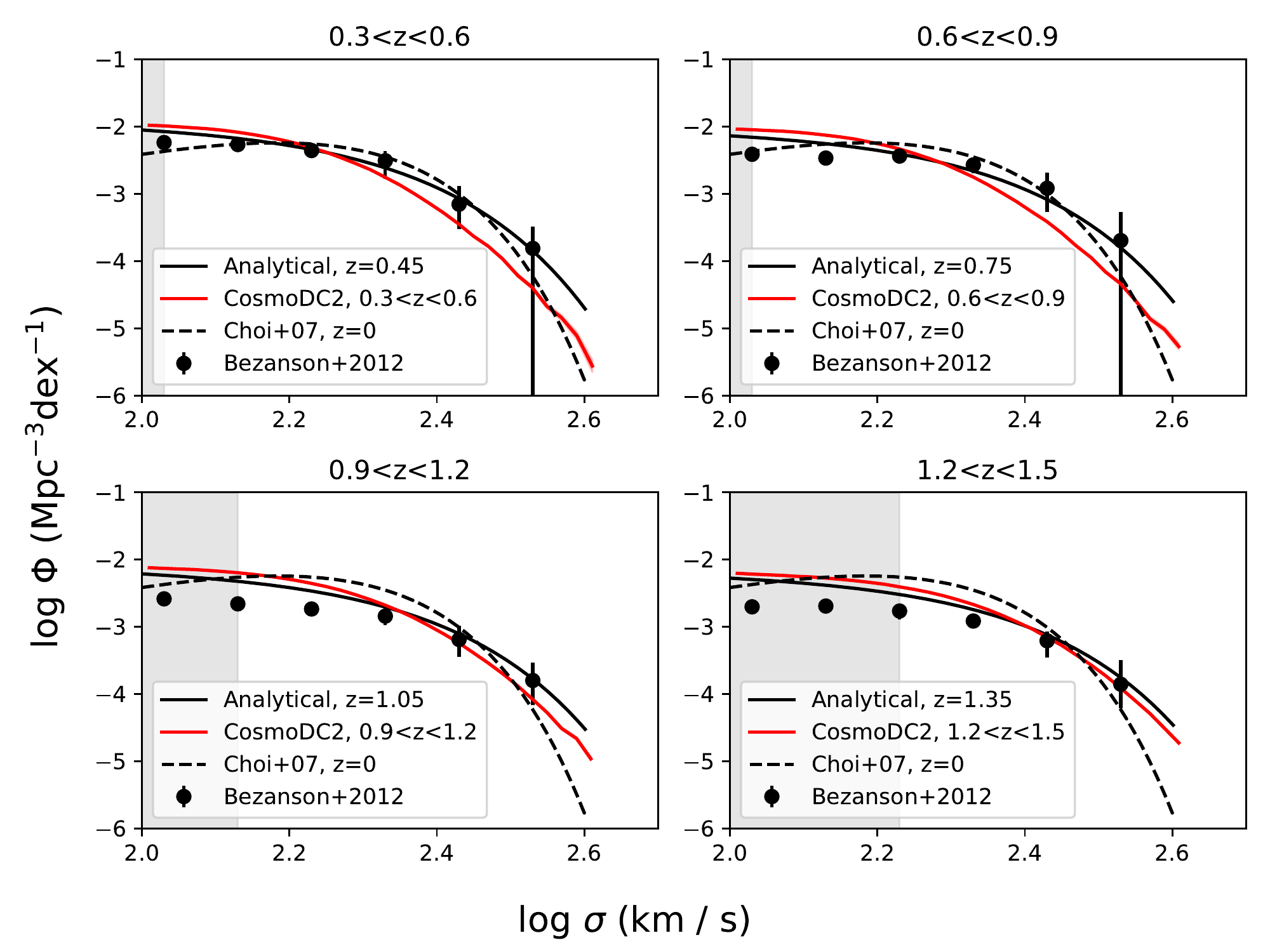}
    \caption{The velocity dispersion functions. {\em Left:} The local VDF. 
    The black solid line shows the analytical VDF adopted by this work, 
    and the red solid line illustrates the VDF of the CosmoDC2 catalog. The open squares and circles are the observed
    VDFs from \citet{sohn17} and \citet{hasan19}. All these VDFs are in good agreement with each other.
    We also include the VDFs used in previous studies of high-redshift lensed quasars for comparison (Section \ref{sec:discussion}).
    {\em Right:} the VDF beyond the local universe. The black dots are the observed VDF from \citet{bezanson12},
    and the gray area marks the mass range where \citet{bezanson12} suggested that their galaxy sample has low completeness.
    Both the analytical VDF and the CosmoDC2 VDF are consistent with the observed ones except at low-mass end,
    which could be a result of the incompleteness of the observed galaxy sample.
    We also include the local VDF from \citet{choi07} (black dashed line)
    which are used in previous studies to model the deflectors.
    }
    \label{fig:vdf}
\end{figure*}

The observed galaxy VDFs are well described by Schechter functions 
\citep[e.g.,][]{choi07,chae10}. In this work, we use the following form of the Schechter function:
\begin{equation}\label{eq:vdf}
    \phi(\sigma,z_d)d\sigma = \phi_* \left(\frac{\sigma}{\sigma_*}\right)^a \exp\left[-\left(\frac{\sigma}{\sigma_*}\right)^b\right]\frac{d\sigma}{\sigma}
\end{equation}
where $\phi(\sigma,z_d)d\sigma$ is the number density of galaxies at redshift $z_d$
which have velocity dispersions within $(\sigma, \sigma+d\sigma)$.
In principle, the parameters $\phi_*, \sigma_*, a$ and $b$ may evolve with redshift.
We adopt the parameterized local VDF from \citet{hasan19} 
and assume the redshift evolution of $\phi_*$ and $\sigma_*$ suggested by \citet{geng21}.
The resulting VDF has 
$\phi_*=6.92\times10^{-3}(1+z_d)^{-1.18}\text{ Mpc}^{-3}$, 
$\sigma_*=172.2\times(1+z_d)^{0.18}\text{ km s}^{-1}$, $a=-0.15$ and $b=2.35$.
In Figure \ref{fig:vdf}, 
we compare our parameterized VDF and the observed values in the local universe from \citet{sohn17} and \citet{hasan19},
and at $0.3<z_d<1.5$ from \citet{bezanson12}.
This parameterized, redshift-evolving VDF is in good agreement with the observations at $z_d<1.5$,
which covers the majority of the deflector population (Section \ref{sec:taum}).

\subsection{Quasar Luminosity Functions}

QLFs are usually parameterized by a broken power-law 
\citep{pei95}, which accurately describes
the QLF at all redshifts \citep[e.g.,][]{kulkarni19}:
\begin{equation}
    \Phi(M, z_s) = \frac{\Phi^*(z_s)}{10^{0.4(\alpha+1)(M-M^*)}+10^{0.4(\beta+1)(M-M^*)}}
\end{equation}
where $\alpha$ and $\beta$ are the faint- and bright-end slopes, $M^*$ is the break magnitude and $\Phi^*$ is the normalization.
For $M$ and $M^*$, we use the absolute magnitude at rest-frame 1450{\AA} $(M_{1450})$ throughout this paper.
We use the median value of $m-M_{1450}$ of $z_s\sim6$ mock quasars (Section \ref{sec:mock}) 
to convert apparent magnitudes $m$ to absolute magnitudes.
Since the shape of QLF determines the magnification bias, 
we explore a wide range of $\alpha,\beta$ and $M^*$ values in Section \ref{sec:magbias}.
The normalization $\Phi^*$ has no impact on the magnification bias and thus the multiply-imaged fraction.

\section{Mock catalog method} \label{sec:mock}

In addition to the analytical model, we also use mock catalogs to investigate the lensed quasar population.
The mock catalogs not only serve as independent tests which validate the analytical model, 
but also illustrate the impact of some simplifications, e.g., using SIS instead of SIE to describe deflectors.
The methods of mock catalog generation is described in Yue et al. (2021, submitted)
and is briefly summarized here.

We use the CosmoDC2 mock galaxy catalog \citep{dc2} to model the deflector galaxy population.
CosmoDC2 is a synthetic mock catalog that covers a sky area of 440 deg$^2$ and out to redshift $z=3$.
We use SIEs to describe the mass profile of the deflector galaxies. 
CosmoDC2 provides the mass, half-light radius and ellipticity of each mock galaxy,
and we estimate the velocity dispersion using a simple relation based on the virial theorem.
The resulting VDF is shwon in Figure \ref{fig:vdf}, 
which is consistent with observations and is in good agreement with the
analytical VDF out to $z_d\sim1.5$.

We then generate two types of mock background sources to study the statistics related to $\tau_m$ and $B$, respectively:
\begin{enumerate}
    \item Sources at fixed redshifts. They are used to model the redshift evolution of $\tau_m$. 
    Since the CosmoDC2 catalog only contains mock galaxies at $z_d<3$,
    it does not provide complete descriptions of the deflectors for sources at $z_s>3$.
    As such, we generate three sets of mock sources at $z_s=1, 2$ and 3.
    \item Simulated quasars at $5.5<z_s<6.5$, which are used to investigate the impact of QLF parameters on the
    magnification bias, $B$, for quasars at $z_s\sim6$. We generate these simulated quasars using SIMQSO \citep{mcgreer13}.
    SIMQSO generates mock quasar catalogs with simulated spectra according to the input QLF.
    Specifically, we consider two sets of QLFs: 
     the ``steep" set which has $\alpha=-2, M^*=-27$ \citep[e.g.,][]{mcgreer18}, 
     and the ``shallow" set which has $\alpha=-1.3, M^*=-25$ \citep[e.g.,][]{matsuoka18}.
    For each set, we consider a range of bright-end slopes at $-3.4\le\beta\le-2.6$. 
    We adopt the normalization $\Phi^*$ from \citet{matsuoka18}. 
\end{enumerate}
We assign random positions to the sources in the CosmoDC2 field, 
and use \texttt{GLAFIC} \citep{glafic} to perform lens modeling and identify multiply-imaged systems. 
For the second type of sources (i.e., mock quasars),
we run multiple random realizations to suppress the Poisson noise,
such that the number of lensed quasars are equivalent to $20\times$ sky area.

It is useful to keep in mind the differences between the mock catalog and the analytical model.
The VDFs of the two approaches are close but have small differences, 
and we use SIE instead of SIS to model the deflectors for mock catalogs. 
Besides, the mock quasars generated by SIMQSO have a variety of continuum slopes and emission line properties
to match the observed distribution,
while in the analytical model, we perform a simple $k-$correction to convert absolute to apparent magnitudes.
As a result, we do not expect that the two methods should give the same result.
Nonetheless, as we will show in Section \ref{sec:stat}, the predictions of the two approaches are largely consistent,
which suggests that both approaches are solid.

\section{Results} \label{sec:stat}

We focus on the multiply-imaged fraction $F_\text{multi}$ of $z_s\sim6$ quasars in this work.
According to Equation \ref{eq:fmulti}, $F_\text{multi}$ is a function of the lensing optical depth $\tau_m$
and the magnification bias $B$. We describe the two factors respectively in the following subsections.

\subsection{Lensing Optical Depth} \label{sec:taum}

We first present the lensing optical depth, $\tau_m$, predicted by the analytical model and the mock catalog.
Since the CosmoDC2 catalog does not have galaxies at $z_d>3$, we only use the mock catalog to calculate the lensing optical depth
for sources at $z_s\le3$. In general, we validate our methods by comparing the mock catalog and the analytical model
at $z_s\le3$, and make predictions using the analytical model at higher redshifts.

\begin{figure*}
    \centering
    \includegraphics[width=0.48\linewidth]{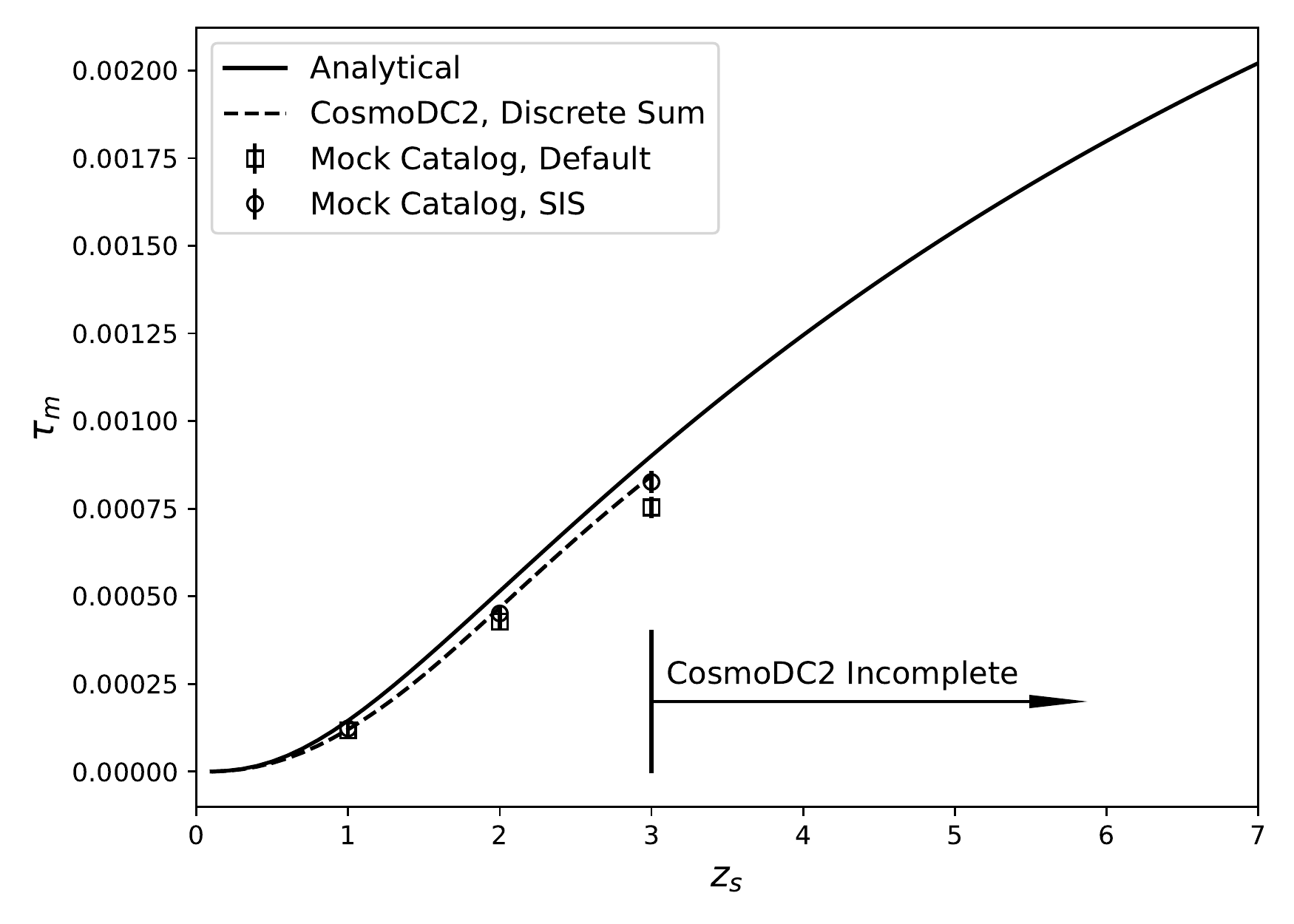}
    \includegraphics[width=0.48\linewidth]{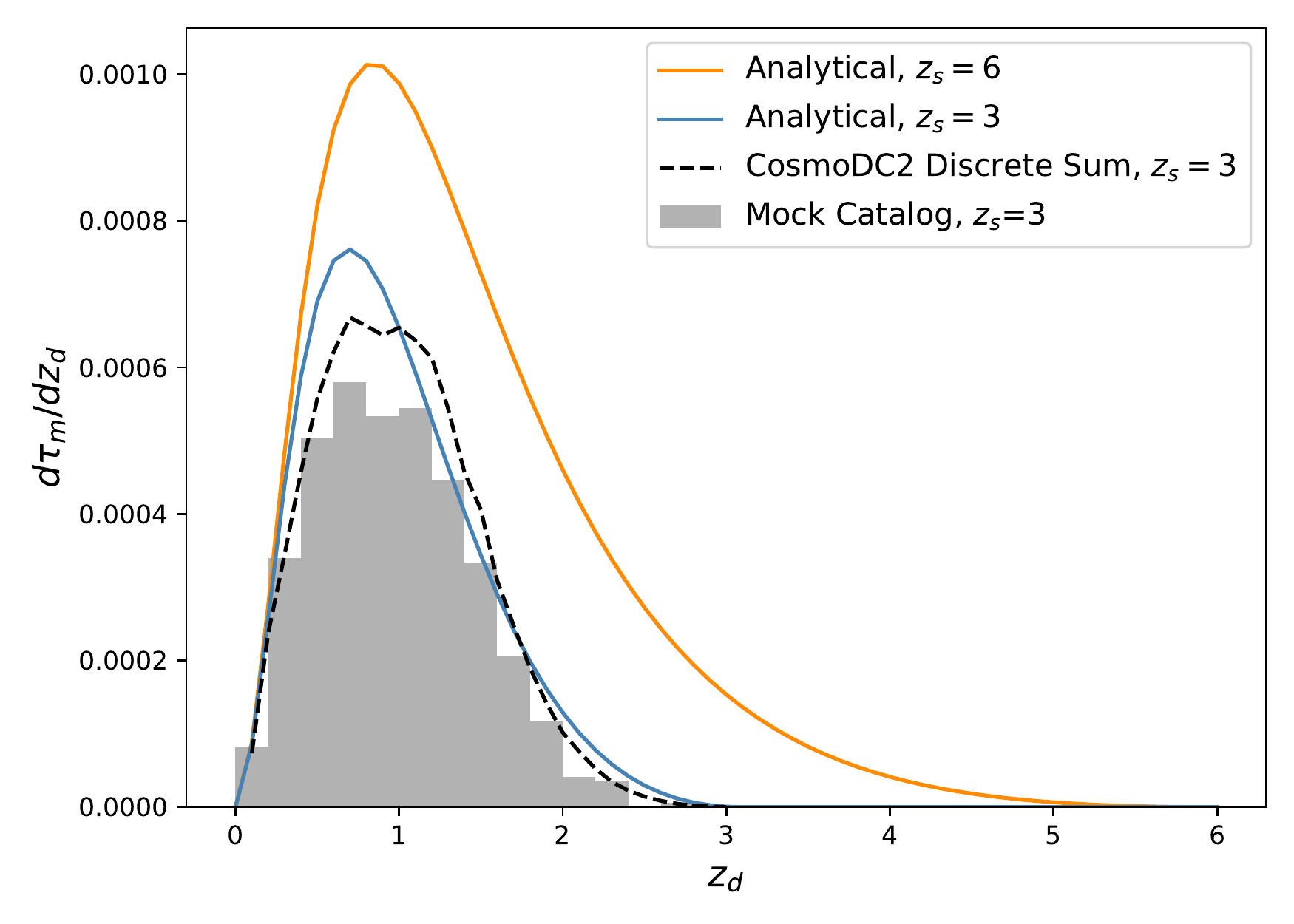}
    \caption{{\em Left:} The lensing optical depth as a function of source redshift. 
    Note that the CosmoDC2 catalog do not contain deflectors at $z_d>3$, so that the mock catalog is incomplete at $z_s>3$.
    The analytical model has $\tau_m=1.8\times10^{-3}$ at $z_s=6$.
    At $z_s<3$, the outputs of the analytical model and the mock catalog are similar.
    We include two additional experiments to validate the results, 
    the CosmoDC2 Discrete Sum and the Mock Catalog SIS, which further demonstrate that our analysis is reliable
    (see text for details).
    {\em Right:} The distribution of deflector redshifts. For $z_s=6$, the analytical model suggests that 
     most of the deflectors have $0.5<z_d<2.5$, and the distribution peaks at $z_d\sim1$. 
     We also include the predictions for $z_s=3$ as a test of our results, 
     for which the analytical method is consistent with the mock catalog.
    }
    \label{fig:taum}
\end{figure*}

The left panel of Figure \ref{fig:taum} illustrates the predicted lensing optical depth as a function of source redshift $z_s$.
At $z_s\le3$, the outputs of the analytical model and the mock catalog are similar.
This comparison suggests that both the analytical method and the mock catalog method are reliable.
We use the analytical model to predict the lensing optical depth at $z_s>3$,
which gives $\tau_m=1.8\times10^{-3}$ for $z_s=6$.

To further validate our analysis, we include two additional experiments:
the ``CosmoDC2 Discrete Sum" where we change the integration in Equation \ref{eq:tm} to discrete sums and calculate $\tau_m$
using the CosmoDC2 VDF, and the ``Mock Catalog SIS" where we use SIS instead of SIE deflectors when making the mock catalogs.
In other words, the two experiments calculate $\tau_m$ using the analytical method and the mock catalog method 
assuming the same VDF and defletcor mass profile.
Figure \ref{fig:taum} shows that the two experiments provide the same result,
 further proving that both methods are reliable.
 The comparison between the default mock catalogs and the Mock Catalog SIS suggests that
using SIS instead of SIE will overestimate the lensing optical depth by $\sim10\%$.
This is consistent with the result in \citet{sie}, who shows that the cross section of an SIE decreases with ellipticity.

In the right panel of Figure \ref{fig:taum}, we show the redshift distribution of the deflector galaxies for sources at $z_s=6$,
predicted by the analytical model.
We also include the predictions for $z_s=3$ from both the analytical model and the mock catalog for comparison.
Again, the two methods give very similar results for $z_s=3$.
For sources at $z_s=6$, the distribution of deflectors peaks at $z_d\sim1$, 
and the majority of deflector galaxies have $0.5\lesssim z_d\lesssim2.5$.
It is thus important to correctly model the redshift evolution of VDF out to $z_d\gtrsim2$
in order to make precise predictions of $\tau_m$ for high-redshift sources.


It is worth noticing that the analytical model 
gives a higher lensing optical depth than the mock catalogs.
The reason is that the analytical VDF is higher than the CosmoDC2 VDF
at $z_d\sim0.6$ (as shown in Figure \ref{fig:vdf}) and $z_d\gtrsim2$.
Since VDFs beyond the local universe has yet been well-constrained
by observations, the predicted lensing optical depth will inevitably has some systematic errors.
we will discuss this point in more details in Section \ref{sec:lensfrac}.

\begin{figure}
    \centering
    \includegraphics[width=0.95\columnwidth]{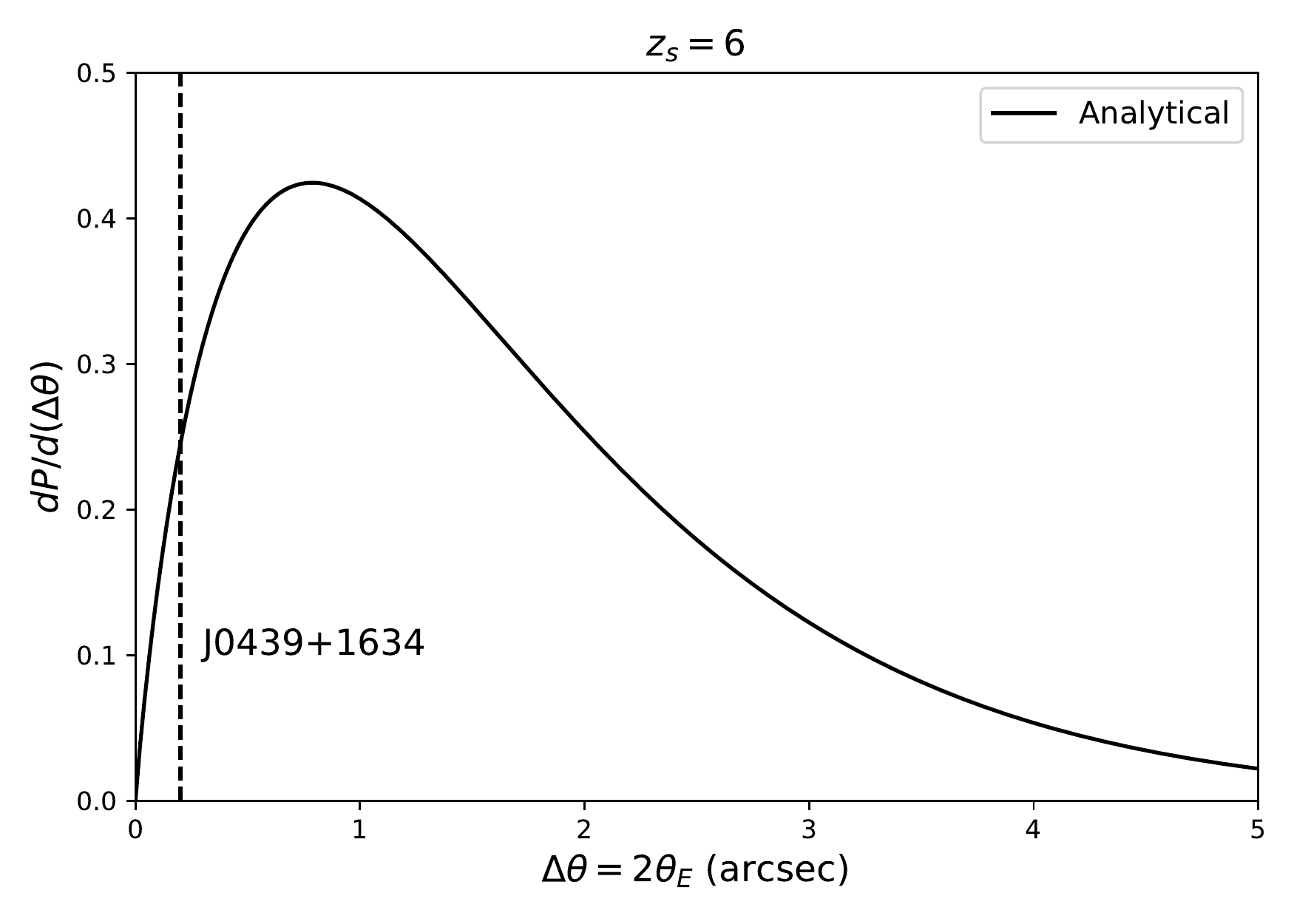}
    \caption{Distribution of the
    lensing separation for sources at $z_s=6$, predicted by the analytical model. About 85\%
    of the lenses have $\Delta \theta>0\farcs5$, and the fraction is $\sim65$\% for $\Delta \theta>1\farcs0$.
    The dashed line shows the separation of J0439+1634 at $z_s=6.52$, which is the only known lensed quasar at $z_s>5$.
    Systems like J0439+1634 are rare but are still possible.}
    \label{fig:sep}
\end{figure}

Figure \ref{fig:sep} shows the distribution of lensing separation for $z_s=6$ sources.
About $65\%~(85\%)$ of the lenses have separation larger than $1''~ (0\farcs5)$.
Most of these lenses will be resolved in surveys such as 
the LSST which has a seeing of $\sim0\farcs7$.
The only lensed quasar at $z>5$ known to date, J0439+1634 at $z=6.52$, is a compact system with $\Delta \theta=0\farcs2$.
Figure \ref{fig:sep} suggests that such systems are rare but possible.

\subsection{Magnification Bias and the Lensed Fraction} \label{sec:magbias}

The magnification bias $B$ is a function of the QLF and $p(\mu)$ (Equation \ref{eq:B}).
We use both the analytical model and the mock catalogs to calculate the magnification bias.
Note that although the mock catalogs do not have deflectors at $z_d>3$,
it still provides correct description for the magnification bias of high-redshift sources, 
since $z_d$ and the VDF do not go into the calculation of $B$.

\begin{figure*}
    \centering    
    \includegraphics[width=0.4\linewidth]{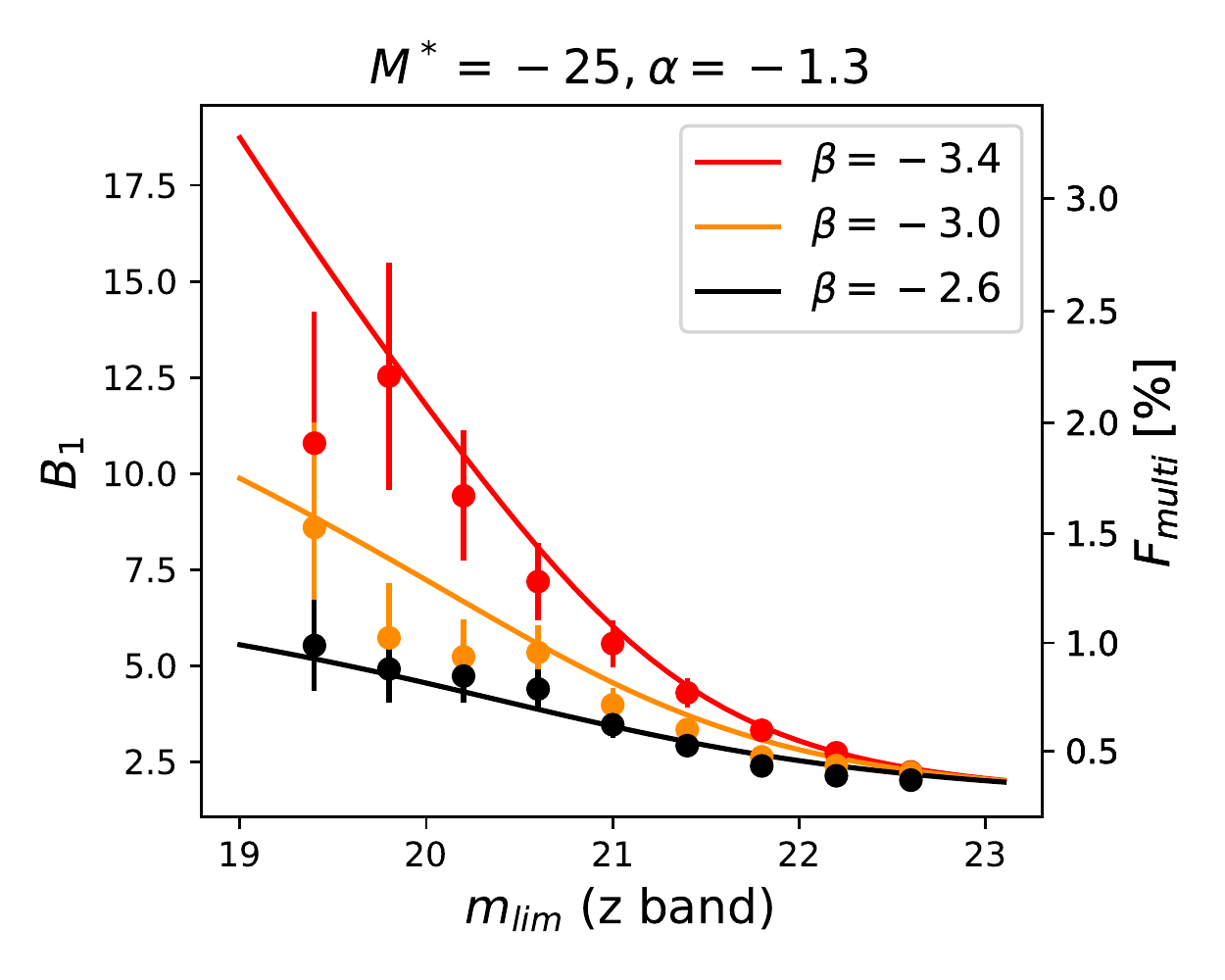}
    \includegraphics[width=0.4\linewidth]{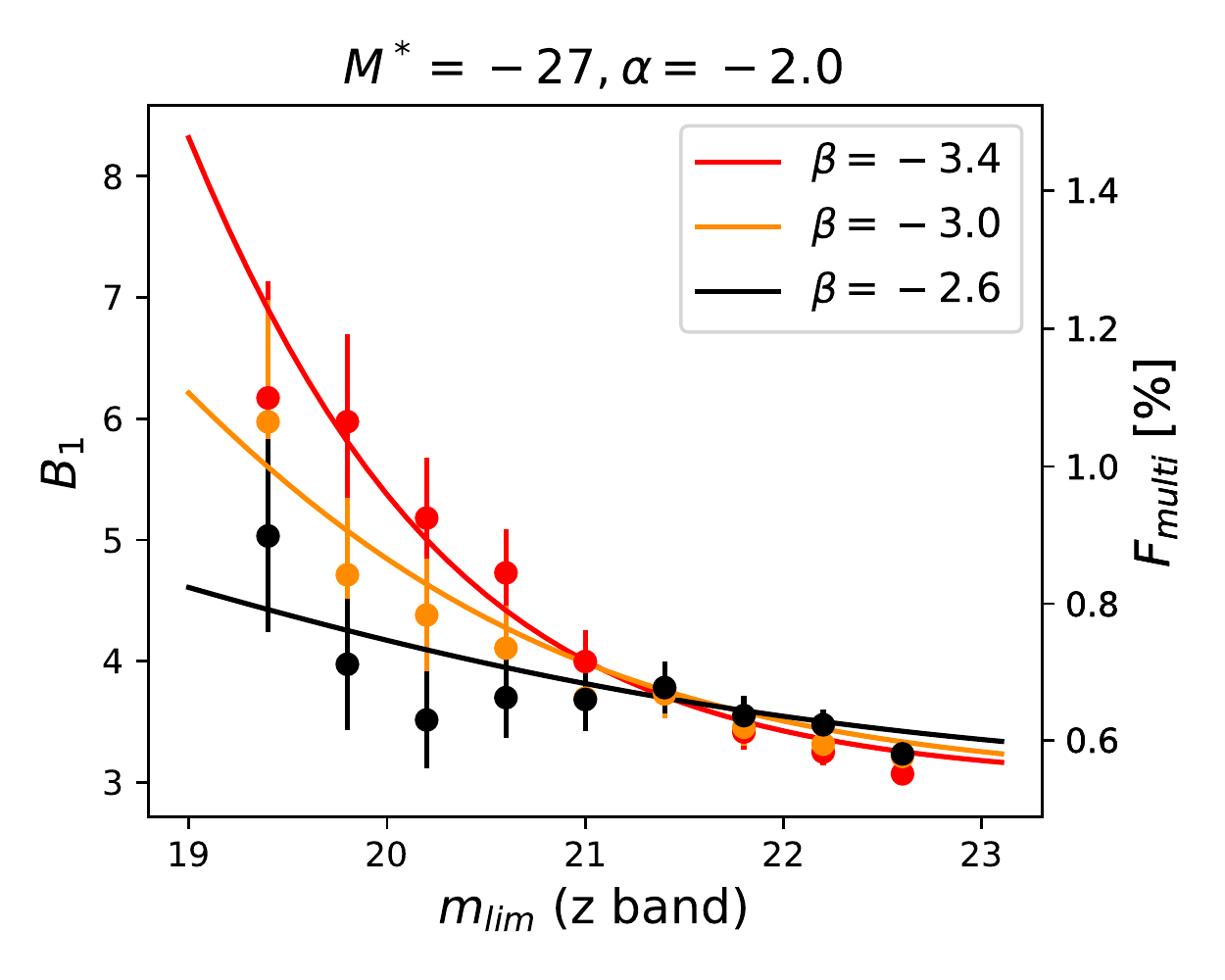}
    \includegraphics[width=0.4\linewidth]{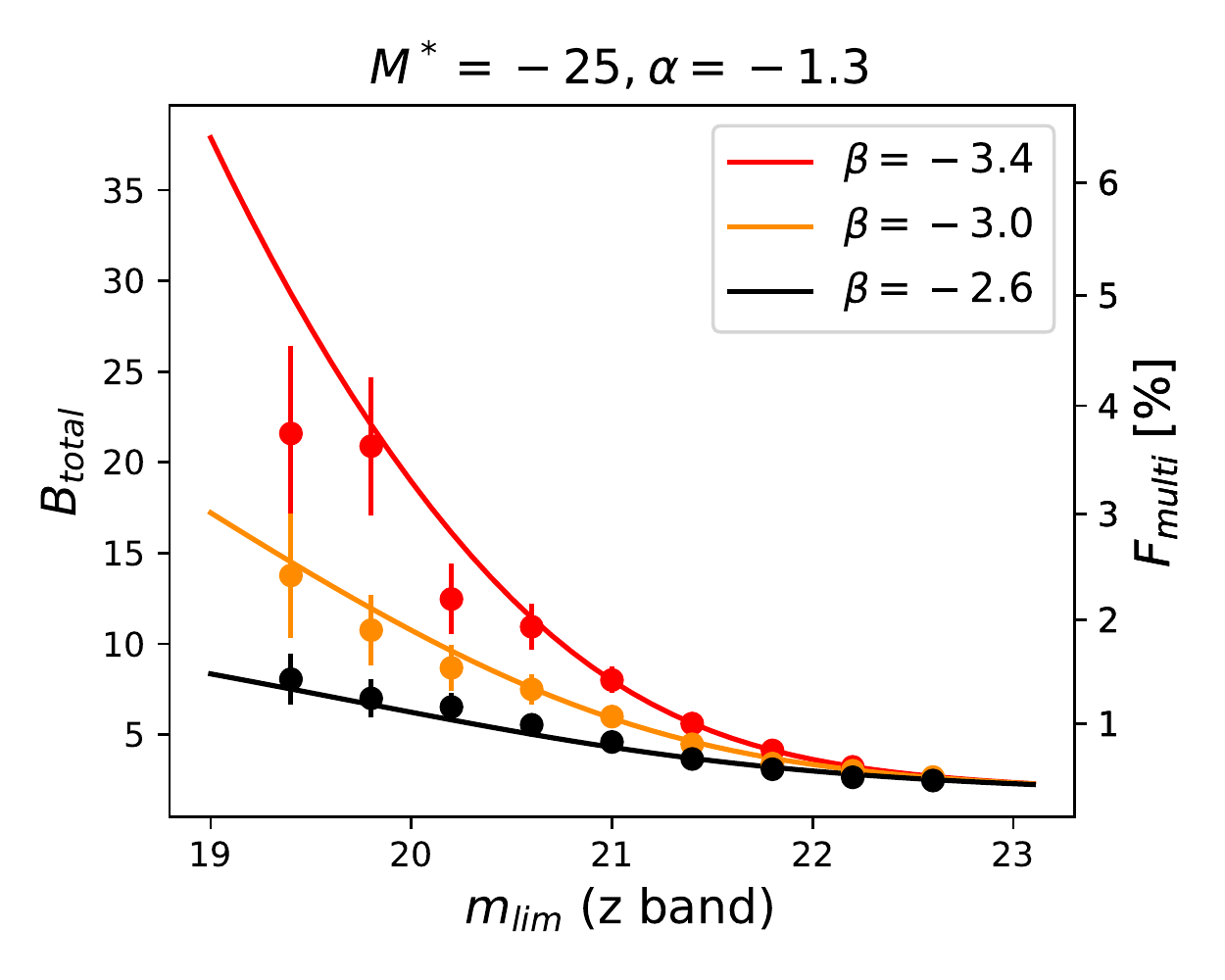}
    \includegraphics[width=0.4\linewidth]{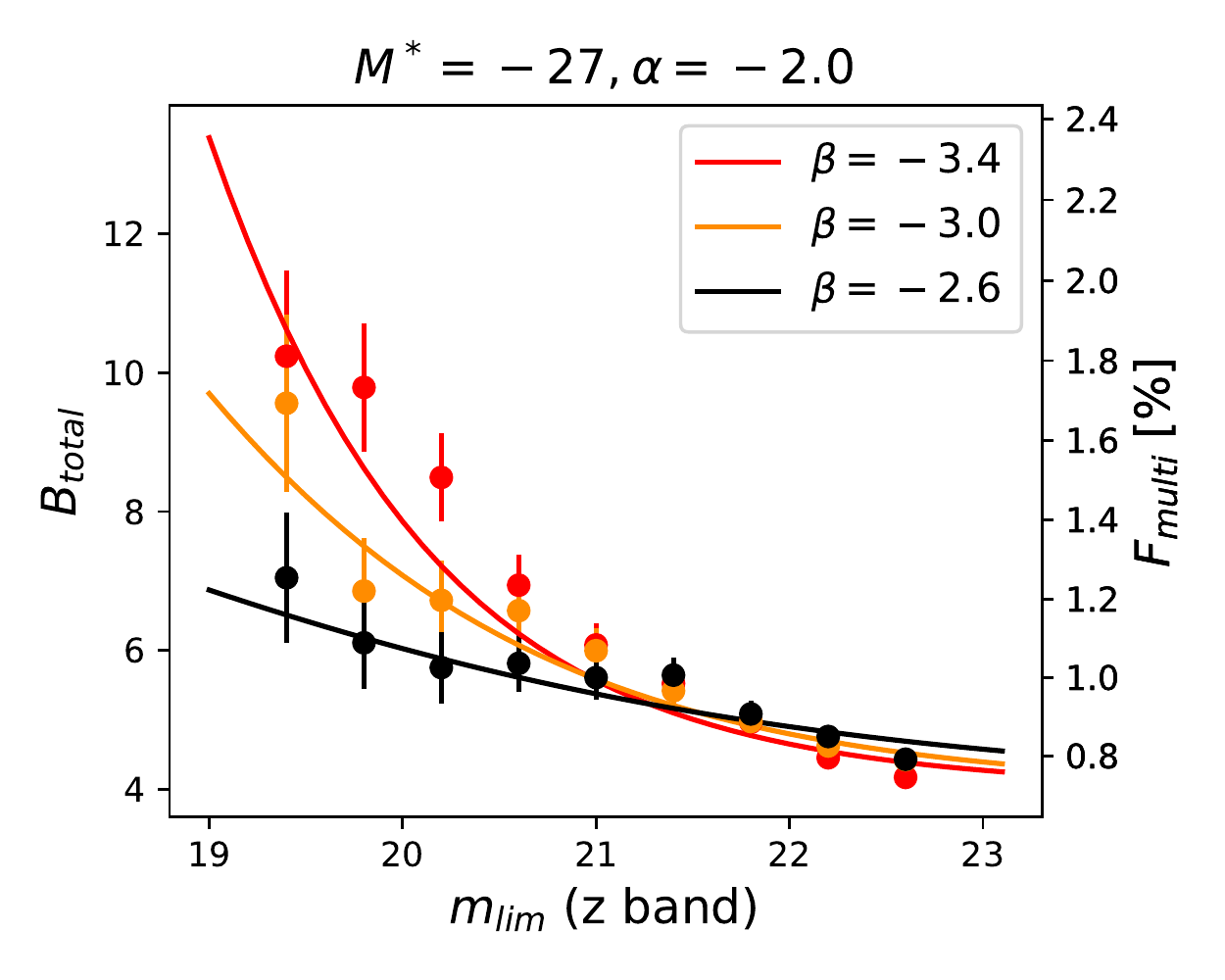}
    \caption{The magnification bias of $z_s\sim6$ lensed quasars as a function of QLF parameters.
    We illustrate two cases, where the top panel shows the magnification bias for the brightest lensed image,
    and the bottom panel is for the total flux of all lensed images.
    In all cases, the analytical model (solid lines) agrees well with the mock catalog (dots with error bars),
    suggesting that the difference between SIS and SIE deflectors is small.
    We assume Poisson statistics for the mock catalog (note that the mock catalogs are equivalent to $20\times$ sky area). 
    The secondary $y-$axis marks the multiply-imaged fraction assuming $\tau_m=1.8\times10^{-3}$.
    The magnification bias is higher for steeper faint-end slope and drops with survey depth.
    For the depth of current surveys of high-redshift quasars $(m_{\text{lim}}\sim22)$, the lensed fraction is
    $F_\text{multi}\sim0.4\%-0.8\%$.}
    \label{fig:magbias}
\end{figure*}

Figure \ref{fig:magbias} shows the predicted magnification bias as a function of QLF parameters. 
We consider two specific cases, where the top panel shows the magnification bias for the brightest lensed image $(B_1)$
and the lower panel shows the bias for the total magnification $(B_\text{total})$.
According to Figure \ref{fig:magbias}, the magnification biases of the two cases differ by $\sim50\%$.
Observationally, these two cases correspond to the high and low spatial resolution limits, respectively,
and real surveys lie between the two special cases.

The predictions of the analytical model and the mock catalogs are consistent in all cases.
This comparison suggests that using SIS in the analytical model gives a good approximation of the real case
where deflectors are elliptical.
The magnification bias is higher for steeper bright-end slope and decrease with the survey depth,
which is a result of the flatter faint-end slope compared to the bright-end.
To calculate the lensed fraction $F_\text{multi}$, we adopt $\tau_m=1.8\times10^{-3}$ for $z_s=6$.
High-redshift quasar searches in the SDSS
 have a depth of $m_{\text{lim}}=20.2$ \citep[e.g.,][]{richards02},
for which the lensed fractions is $F_\text{multi}\sim1\%-3\%$. 
Present-day surveys for $z\sim6$ quasars usually have $m_{\text{lim}}\gtrsim22$ \citep[e.g.,][]{jiang16,matsuoka18},
which corresponds to $F_\text{multi}\sim0.4\%-0.8\%$.
The lensed fraction of the brightest $z_s\sim6$ quasar population $(m_z\lesssim19)$
can be as high as $\sim6\%$ for the steep bright-end slopes.

\begin{figure*}
    \centering    
    \includegraphics[width=0.45\linewidth]{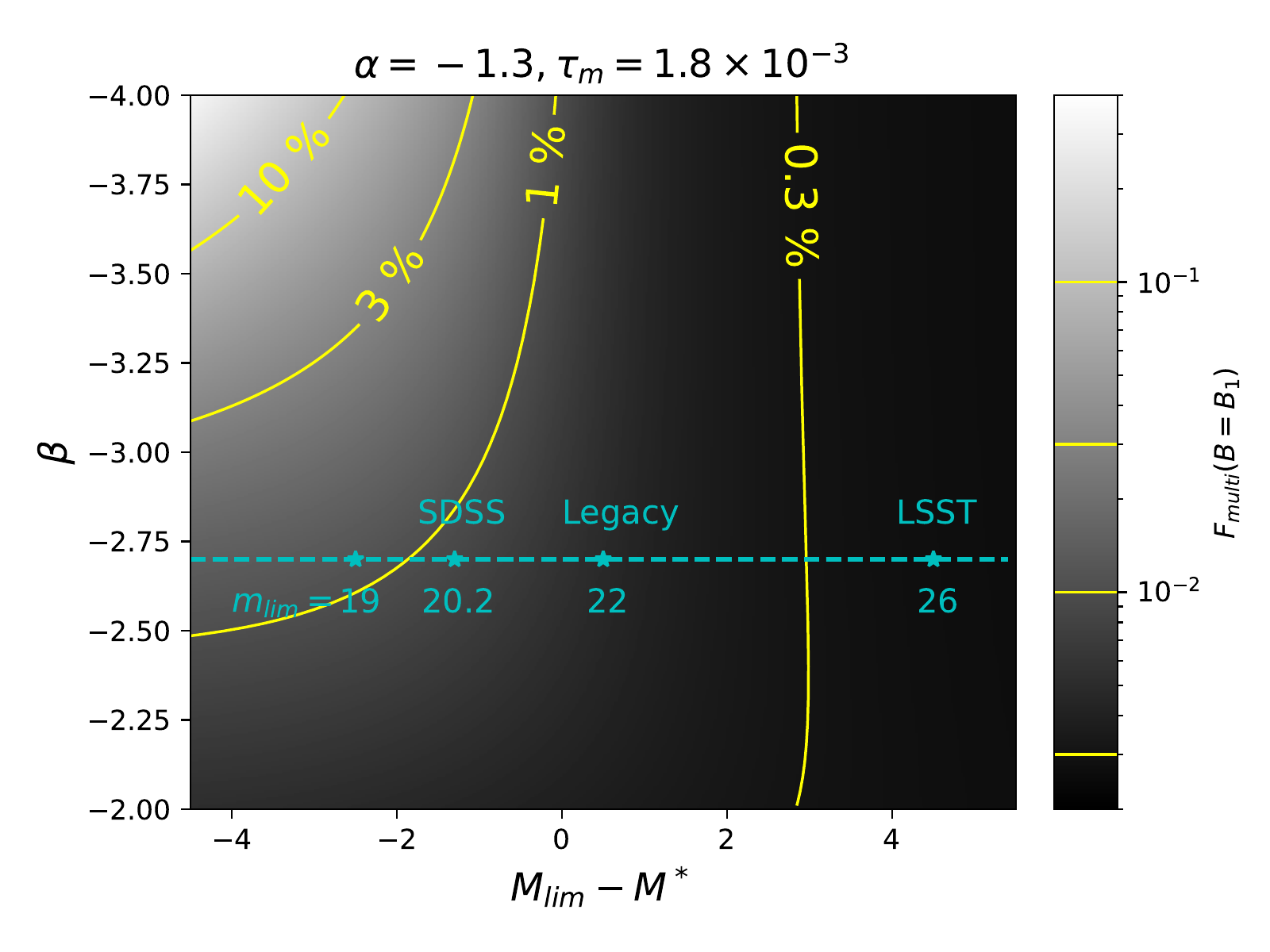}
    \includegraphics[width=0.45\linewidth]{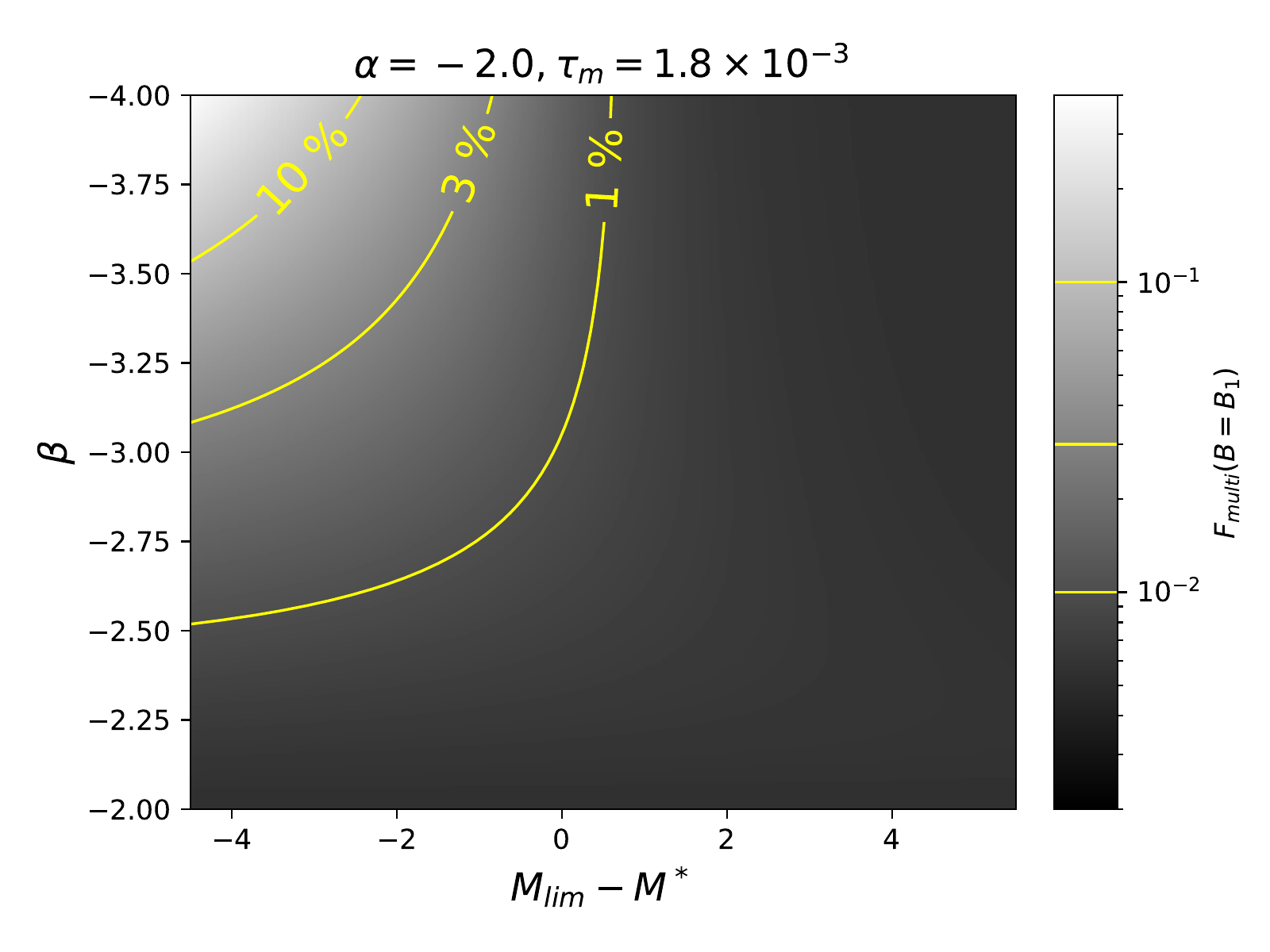}
    \includegraphics[width=0.45\linewidth]{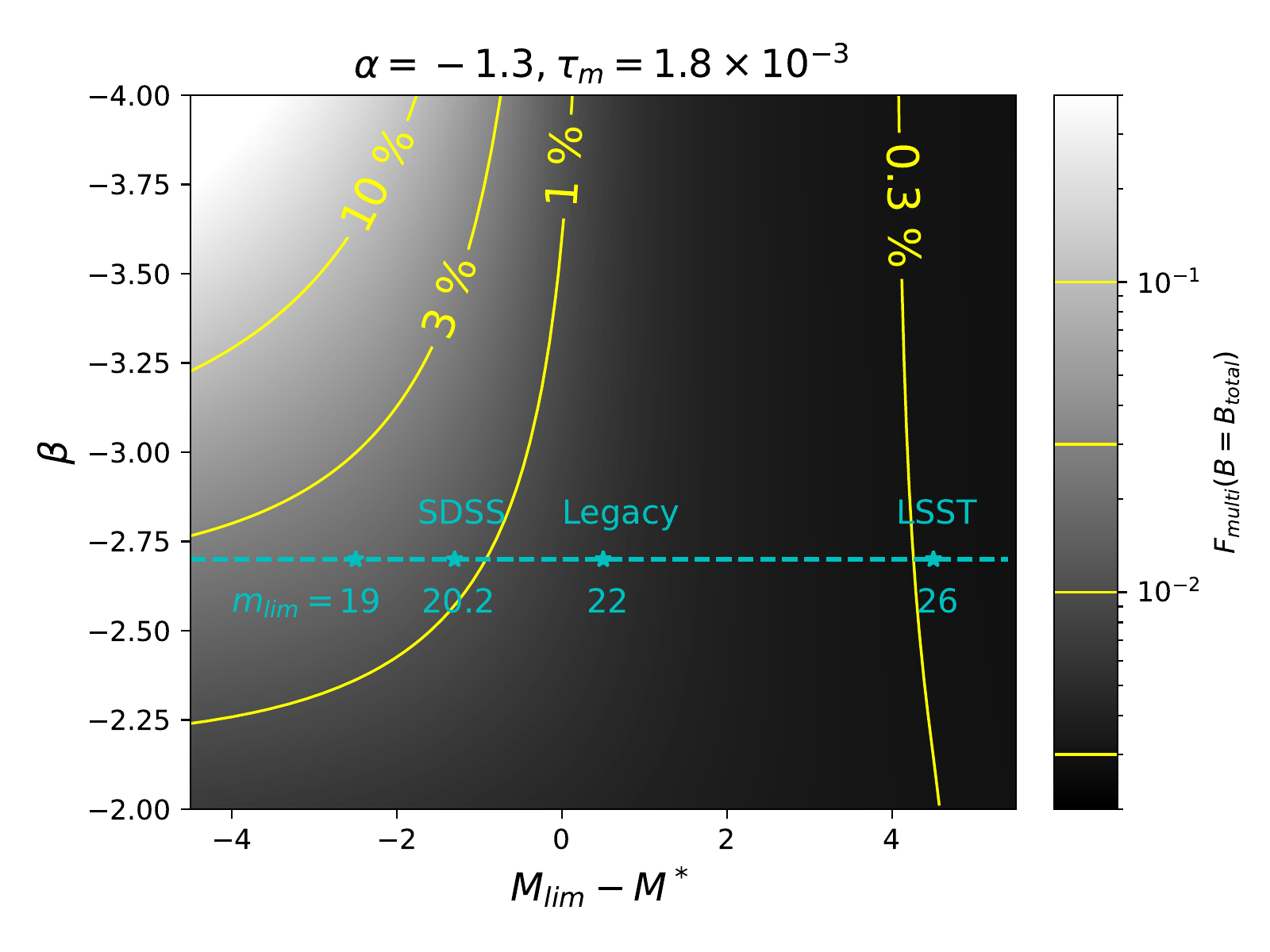}
    \includegraphics[width=0.45\linewidth]{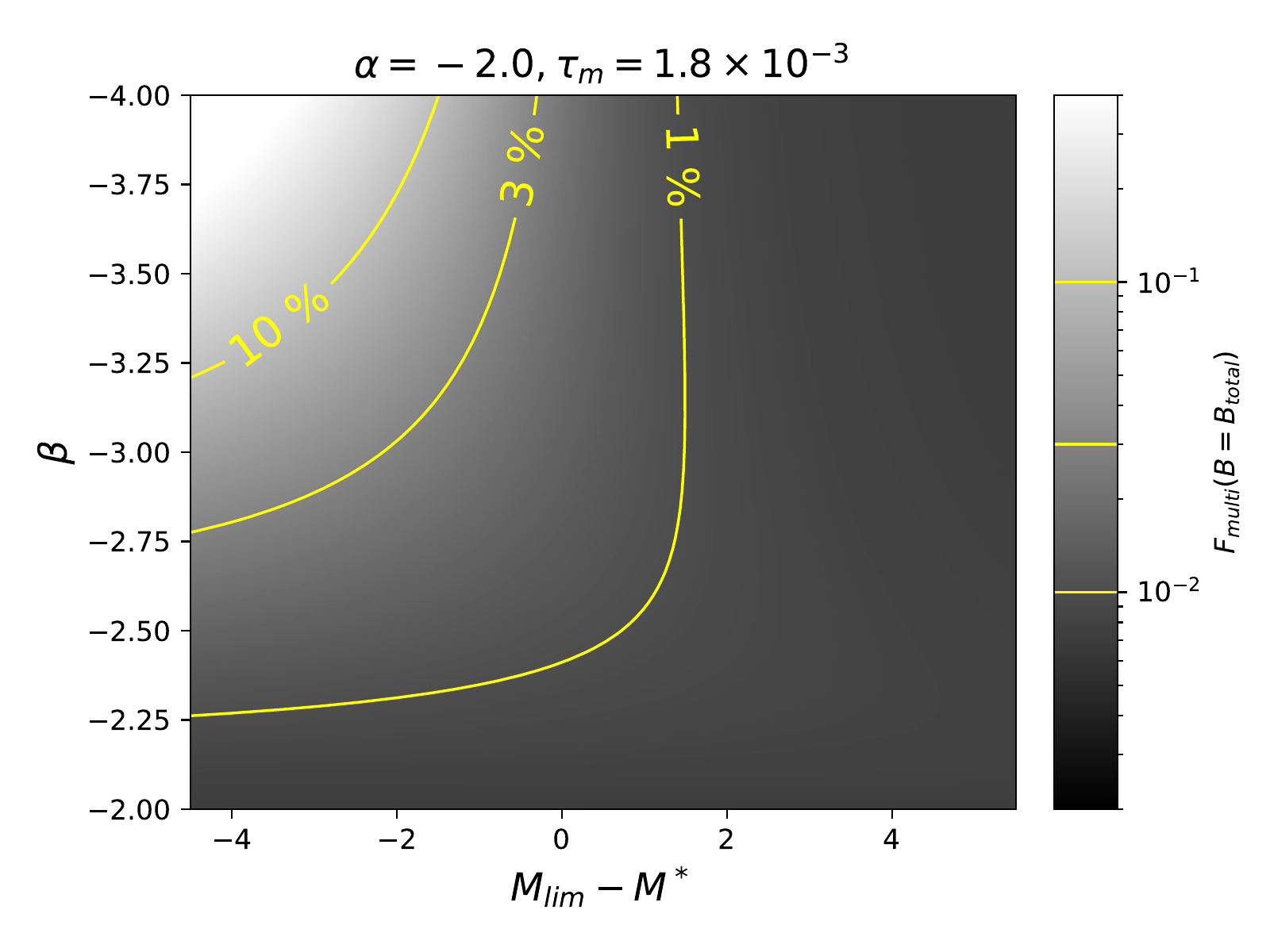}
    \caption{Lensed fraction of $z_s\sim6$ quasars as a function of QLF parameters, 
    predicted by the analytical model. Similar to Figure \ref{fig:magbias}, we assume $\tau_m=1.8\times10^{-3}$, 
    and illustrate the two cases that correspond to the brightest image (top) and the total flux (bottom). 
    The contours mark levels of $10\%, 3\%, 1\%$ and $0.3\%$,
    and these levels are also marked in the colorbars by 
    yellow lines.
    For shallow surveys $(M_{\text{lim}}-M^*\lesssim-2)$ and steep bright-end slopes $(\beta\lesssim-3.5)$,
    the lensed fraction can be as high as $F_\text{multi}\gtrsim10\%$. The fraction drops quickly towards 
    other regions in the parameter space. 
    The cyan dashed lines and asterisks highlight
    the fiducial QLF with $M^*=-25, \alpha=-1.3$ and $\beta=-2.7$,
    which has $F_\text{multi}\sim1\%-2\%$ for the brightest quasars and $F_\text{multi}\sim0.4\%-0.8\%$
    for the depths of current surveys.}
    \label{fig:fmult}
\end{figure*}

In Figure \ref{fig:fmult}, we further explore the impact of QLF parameters on the lensed fraction using the analytical model.
Note that an absolute magnitude of $-25$ corresponds to
an apparent magnitude of $21.5$ at $z_s=6$ under our assumed $k-$correction. 
The multiply-imaged fraction can be as high as $\gtrsim10\%$ 
at the corner of $M_{\text{lim}}-M^*\lesssim-2$ and $\beta\lesssim-3.5$.
The lensed fraction drops quickly to other regions. 
In particular, we highlight the predictions for our fiducial QLF of 
$\alpha=-1.3, \beta=-2.75$ and $M^*=-25$, which are close to the measurements in \citet{matsuoka18} and
\citet{wang19} for $z_s\gtrsim6$ quasars. 
This fiducial QLF gives $F_\text{multi}\sim1\%-2\%$ for the brightest quasars $(m_z\lesssim19)$
and $F_\text{multi}\sim0.4\%-0.8\%$ for a survey depth of $m_z\approx22$.


\section{Discussion} \label{sec:discussion}

\subsection{The Lensed Fraction of High-redshift Quasars} \label{sec:lensfrac}

We can now investigate the discrepancy between observations and theoretical models
in the lensed fraction of high-redshift quasars.
According to Equation \ref{eq:fmulti}, the multiply-imaged fraction $F_\text{multi}$ is a function of
the lensing optical depth $\tau_m$ and the magnification bias $B$.
We first consider the lensing optical depth $\tau_m$, which is determined by the deflector VDF.
Figure \ref{fig:vdf} summarizes the VDFs adopted by previous models of high-redshift lensed quasars.
In general, two approaches have been adopted to model the deflector VDF:

(1) Convert galaxy luminosity functions to VDFs using empirical relations such as 
the Faber-Jackson relation \citep[FJR,][]{fj}. This method was used in \citet{wyithe02} and \citet{pacucci19}. 
Figure \ref{fig:vdf} shows that the VDFs in these two studies are close to each other, and both of them
 are significantly higher than the observed VDFs.
Correspondingly, 
both studies predicted a lensed fraction that is several times higher than our model ($F_\text{multi}\gtrsim5\%$ for $m_\text{lim}=20.2$).
The reason why FJR leads to a high VDF is complicated. 
First, the FJR has a large scatter \citep[$\gtrsim10\text{ km s}^{-1}$, e.g., ][]{bn15}.
Applying a single FJR without scatters will not accurately reproduce the VDF.
Second, the FJR are usually calibrated for massive galaxies with $\log \sigma\gtrsim2.3$ 
\citep[e.g.,][]{forcardi12},
which might not describe less massive galaxies correctly. 

(2) Directly adopt the VDF from galaxy surveys. 
Most previous studies of this kind adopted the  local VDF in \citet{choi07}
and assumed no redshift evolution. For example, \citet{wyithe11} calculated the lensing optical depth 
out to $z_s\sim11$ and gave $\tau_m=3\times10^{-3}$ for $z_s=6$.
Recent observations suggested that the number density of massive galaxies decreases towards high redshifts 
\citep[e.g.,][]{chae10, bezanson12},
indicating that a non-evolving VDF overestimates the lensing optical depth.
As shown in Figure \ref{fig:vdf},
at $z_d\gtrsim0.9$, the \citet{choi07} VDF is higher than the observations and the VDFs in this study,
which explains the difference in the predicted $\tau_m$.
Deflectors at this redshift range dominate the lensing optical depth (Figure \ref{fig:taum}).
This comparison illustrates the importance of using a redshift-evolving VDF.


In addition to using VDFs to describe the deflector population,
some studies also use cosmological simulations to calculate the lensing optical depth.
For example, \citet{hilbert08} used the matter distribution from the Millennium simulation \citep{springel05}
and performed ray-tracing to model the strong-lensing statistics. 
They predict a lensing optical depth that is comparable to our results ($\tau_m\sim1.2\times10^{-3}$ at $z_s=5.7$).

To summarize, it is critical to have an accurate description of the deflector population
when modeling the lensing statistics.
The VDFs adopted by previous studies may have significant uncertainties (usually overestimated),
which could bias the predicted lensing optical depth by a factor of $\sim2-4$.
By applying recent measurements of VDFs, we reach a lower lensing optical depth and thus lower lensed fraction,
which relieves the tension between the theoretical models and the observations.
Since the VDF at $z_d\gtrsim1.5$ are still largely unconstrained, the predicted $\tau_m$ in our model
also have some systematic errors. We estimate this uncertainty to be $\sim30\%$ by comparing the 
analytical VDF and the CosmoDC2 VDF, which are in good agreement at $z_d\lesssim1.5$ 
but show more differences at higher redshifts.
After taking the uncertainties into account, 
the difference between the observed lensed fraction $(F_\text{multi}\approx0.2\%)$ and our model prediction
$(F_\text{multi}\approx0.4\%-0.8\%)$ is marginal.

We then consider the magnification bias $B$.
Previous studies explored a wide range of QLF parameters, 
and the results are similar to this study. This is largely due to that SIS accurately describes the magnification distribution,
$p(\mu)$, as suggested by Figure \ref{fig:magbias}.
However, it is worth noticing that most of the previous studies used the
SDSS quasar survey depth $(m_{\text{lim}}=20.2)$ when predicting the lensed fraction. 
Present-day high-redshift quasar surveys have depths of $m_{\text{lim}}\gtrsim22$,
which lead to a lower expected lensed fraction than the SDSS depth.

\subsection{The Number of Detectable High-Redshift Lensed Quasars in Current and Future Sky Surveys}

Large and deep imaging surveys (e.g., the LSST survey and the Euclid survey \citep[][]{euclid})
will greatly enhance our ability of finding high-redshift quasars.
In this Section, we estimate the number of high-redshift lensed quasars 
that can be detected in current and future sky surveys.

We use the analytical model 
to calculate the number of lensed quasars that can be detected 
given a certain survey depth.
\footnote{The code for the analytical model is available at https://github.com/yuemh/lensQSOsim/tree/main/analytical}
Specifically, we estimate the number of quasars that are intrinsically 
brighter than the survey limit using the 
 QLF from \citet{matsuoka18}, then calculate the 
 number of detectable strongly-lensed quasars using Equation \ref{eq:fmulti}.
 To convert apparent magnitudes to absolute magnitudes for quasars at redshift $z_s$, 
 we generate mock quasars at $z_s-0.1<z_\text{mock}<z_s+0.1$
 following the method in Section \ref{sec:mock},
 and adopt the median value of $m-M_{1450}$ of these mock quasars.
Since the majority of the $z_s\sim6$ lenses will be resolved
in upcoming sky surveys that have resolution of $\lesssim0\farcs7$,
we use the magnification bias for the brightest image, $B=B_1$.

\begin{figure} 
    \centering    
    \includegraphics[width=0.95\linewidth]{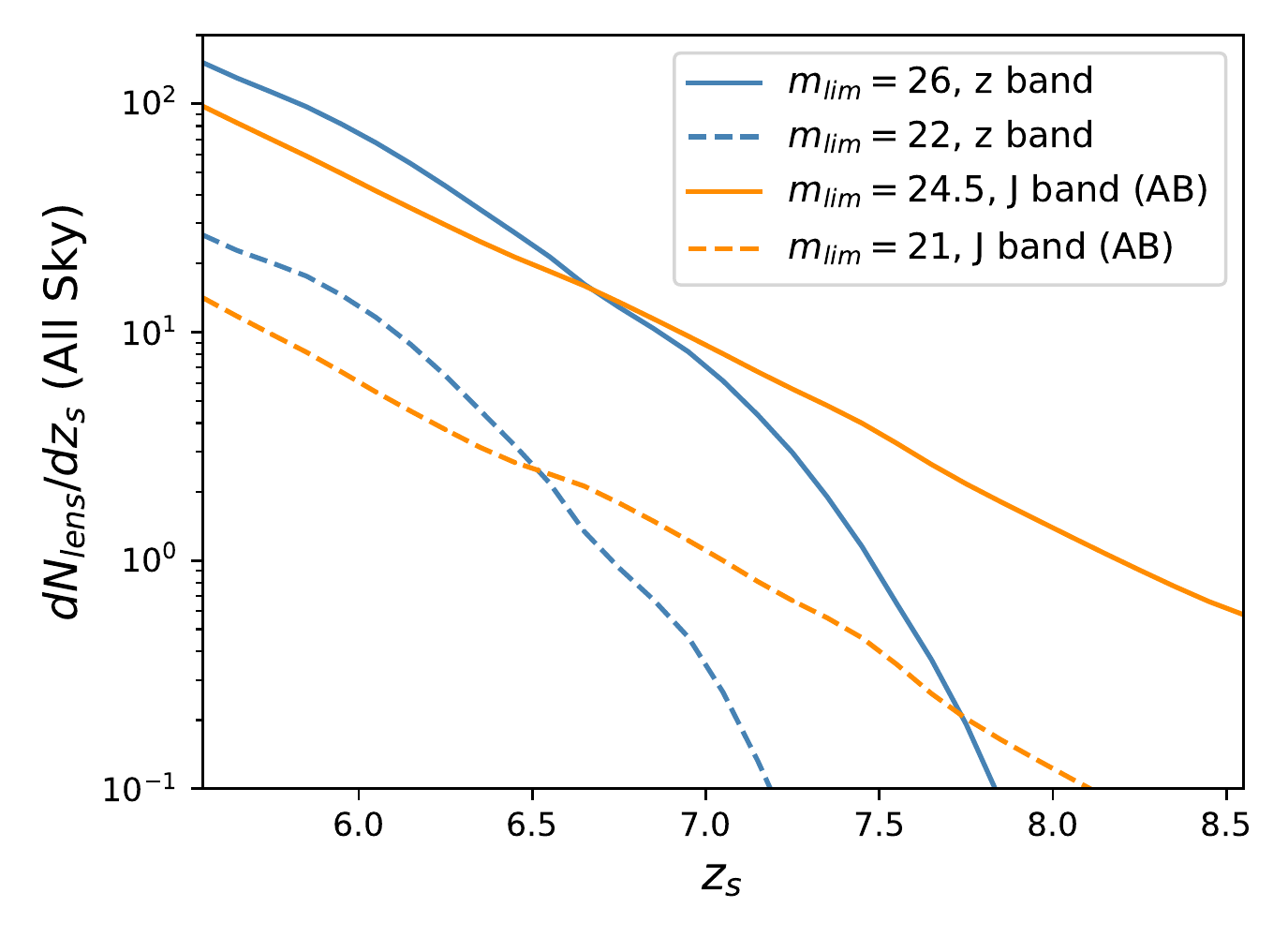}
    \caption{Numbers of detectable lensed quasars as a function of survey depth and quasar redshift.
    The solid lines correspond to the depths of the LSST survey $(m_z\approx22)$ 
    and the {\em Euclid} survey $(m_{J,\text{AB}}\approx24.5)$,
    and the dashed lines are depths of current imaging surveys (see text for details).
    Current sky surveys (optical + NIR) can detect $\sim15$ lensed quasars at $z_s>5.5$ over the whole sky.
    The LSST survey can detect $\sim45$ lensed quasars at $z_s>5.5$ in its footprint,
    and the {\em Euclid} survey can detect $\sim4$ at $z_s>6.5$.}
    \label{fig:dndz}
\end{figure}

Figure \ref{fig:dndz} presents the predicted number density of 
lensed quasars that have the brightest lensed image detectable 
as a function of survey depth and redshift.
We first consider optical imaging surveys, 
for which we use their $z-$band depths to compute the number of detectable lensed quasars.
Current optical surveys have a typical depth of $m_z=22$ 
\citep[e.g., the DESI Legacy Imaging Survey,][]{legacy},
yielding $14$ detectable lensed quasars at $z_s>5.5$ over the whole sky.
The LSST survey will reach a depth of $m_z\approx26$,
with which we can detect 89 lensed quasars at $z_s>5.5$ over the whole sky
($\sim45$ over the $20,000\text{ deg}^2$ LSST footprint).

For quasars at $z_s\gtrsim6.5$, the Ly$\alpha$ wavelength is redshifted to $\lambda_\text{obs}\gtrsim9120\text{\AA}$,
and a substantial fraction of flux in $z-$band is absorbed by
 the highly neutral intergalactic medium.
 Consequently, the numbers of detectable lensed quasars in optical surveys 
 drop quickly at this redshift range.
 We thus consider the expected outputs of near-infrared (NIR) imaging surveys.
 Present-day NIR surveys like the UKIRT Hemisphere Survey \citep[UHS,][]{uhs} 
 and the VISTA Hemisphere Survey \citep[VHS,][]{vhs} have depths of
 $m_J\approx21$ (AB magnitude), leading to $\sim1-2$ detectable lensed quasars 
 at $z_s>6.5$ over the entire sky.
 In the upcoming {\em Euclid} survey that will reach a depth of $m_J=24.5$,
there will be 12 detectable lensed quasars at $z_s>6.5$ over the sky
($\sim4$ over the $14,000\text{ deg}^2$ {\em Euclid} footprint).

To summarize, current sky surveys can only detect about 15 lensed quasars at $z_s>5.5$
over the entire sky. Next-generation sky surveys are needed to build the 
first large sample of high-redshift lensed quasars. Besides,
our model predicts that the {\em Euclid} survey can only detect $\sim0.6$ lensed quasars at $z_s>7.5$,
meaning that we are not likely to find a lensed quasar beyond this redshift in the near future.

Note that the numbers quoted above correspond to {\em detectable} lensed quasars; 
the outputs of real surveys heavily depend on the completeness of 
candidate selection methods and can be exceedingly complicated 
(see related discussion in Yue et al. 2021, submitted). 
If we only count the area with high galactic latitude (e.g., $|b| \gtrsim 30$ deg)
 and assume a high survey completeness (e.g., $\gtrsim80\%$),
the model suggests that we can only find a handful of lensed quasars at $z_s > 5.5$ in current sky surveys.
The completeness of real surveys could be much lower \citep[e.g.,][]{fan19,pacucci19}. 
We expect that upcoming sky surveys like the LSST will find several tens of lensed quasars at $z_s > 5.5$.

\section{Conclusion} \label{sec:conclusion}

We revisit the models of the high-redshift lensed quasar population, 
focusing on the lensed fraction $F_\text{multi}$ of $z_s\sim6$ quasars.
We adopt recent measurements of VDFs and explore a wide range of QLF parameters,
using both analytical methods and mock catalogs.
Our models suggest a lensing optical depth of $\tau_m=1.8\times10^{-3}$ for sources at $z_s=6$,
and a lensed fraction of $F_\text{multi}\sim 0.4\%-0.8\%$ 
for ordinary QLF parameters and current survey depth $(m_\text{lim}\approx22)$.
For the brightest $z_s\sim6$ quasars $(m_z\lesssim19)$, 
the fraction is $\sim2\%$ for ordinary QLF parameters and
can be as high as $\sim6\%$ for steep QLF bright-end slopes.

By comparing our models to previous studies, 
we illustrate that it is critical to use an accurate, redshift-evolving VDF. 
Inaccurate VDFs will bias the predicted $\tau_m$ by a factor of several.
Adopting VDFs from recent measurements
relieves the tension between the observed lensed fraction and the model predictions.
As the VDF at $z_d\gtrsim2$ is still poorly constrained, 
we estimate that our model still has a systematic uncertainty of $\sim30\%$ for $\tau_m$ and thus $F_\text{multi}$.

Finally, we estimate the number of high-redshift lensed quasars that can get detected in 
present-day and future imaging surveys.
Our model suggests that there are $\sim15$ lensed quasars beyond the magnitude limit of current 
wide-area imaging surveys.
Deeper surveys such as the LSST survey and the {\em Euclid} survey
will find several tens of lensed quasars at $z_s>5.5$
and are necessary for building the first large sample of high-redshift lensed quasars.

\acknowledgments
We appreciate the valuable comments from Dr. Fabio Pacucci, Prof. Stuart Wyithe and the anonymous Referee.
MY, XF and JY acknowledges supports by NSF grants AST 19-08284.
FW thanks the support provided by NASA through the NASA Hubble Fellowship grant \#HST-HF2-51448.001-A awarded by the Space Telescope Science Institute, which is operated by the Association of Universities for Research in Astronomy, Incorporated, under NASA contract NAS5-26555.

\bibliography{sample631}{}

\begin{thebibliography}{}
\expandafter\ifx\csname natexlab\endcsname\relax\def\natexlab#1{#1}\fi
\providecommand{\url}[1]{\href{#1}{#1}}
\providecommand{\dodoi}[1]{doi:~\href{http://doi.org/#1}{\nolinkurl{#1}}}
\providecommand{\doeprint}[1]{\href{http://ascl.net/#1}{\nolinkurl{http://ascl.net/#1}}}
\providecommand{\doarXiv}[1]{\href{https://arxiv.org/abs/#1}{\nolinkurl{https://arxiv.org/abs/#1}}}

\bibitem[{{Barone-Nugent} {et~al.}(2015){Barone-Nugent}, {Wyithe}, {Trenti},
  {Treu}, {Oesch}, {Bouwens}, {Illingworth}, \& {Schmidt}}]{bn15}
{Barone-Nugent}, R.~L., {Wyithe}, J.~S.~B., {Trenti}, M., {et~al.} 2015,
  \mnras, 450, 1224, \dodoi{10.1093/mnras/stv633}

\bibitem[{{Bezanson} {et~al.}(2012){Bezanson}, {van Dokkum}, \&
  {Franx}}]{bezanson12}
{Bezanson}, R., {van Dokkum}, P., \& {Franx}, M. 2012, \apj, 760, 62,
  \dodoi{10.1088/0004-637X/760/1/62}

\bibitem[{{Chae}(2010)}]{chae10}
{Chae}, K.-H. 2010, \mnras, 402, 2031, \dodoi{10.1111/j.1365-2966.2009.16073.x}

\bibitem[{{Choi} {et~al.}(2007){Choi}, {Park}, \& {Vogeley}}]{choi07}
{Choi}, Y.-Y., {Park}, C., \& {Vogeley}, M.~S. 2007, \apj, 658, 884,
  \dodoi{10.1086/511060}

\bibitem[{{Comerford} {et~al.}(2002){Comerford}, {Haiman}, \&
  {Schaye}}]{comerford02}
{Comerford}, J.~M., {Haiman}, Z., \& {Schaye}, J. 2002, \apj, 580, 63,
  \dodoi{10.1086/343116}

\bibitem[{{Dey} {et~al.}(2019){Dey}, {Schlegel}, {Lang}, {Blum}, {Burleigh},
  {Fan}, {Findlay}, {Finkbeiner}, {Herrera}, {Juneau}, {Landriau}, {Levi},
  {McGreer}, {Meisner}, {Myers}, {Moustakas}, {Nugent}, {Patej}, {Schlafly},
  {Walker}, {Valdes}, {Weaver}, {Y{\`e}che}, {Zou}, {Zhou}, {Abareshi},
  {Abbott}, {Abolfathi}, {Aguilera}, {Alam}, {Allen}, {Alvarez}, {Annis},
  {Ansarinejad}, {Aubert}, {Beechert}, {Bell}, {BenZvi}, {Beutler}, {Bielby},
  {Bolton}, {Brice{\~n}o}, {Buckley-Geer}, {Butler}, {Calamida}, {Carlberg},
  {Carter}, {Casas}, {Castander}, {Choi}, {Comparat}, {Cukanovaite}, {Delubac},
  {DeVries}, {Dey}, {Dhungana}, {Dickinson}, {Ding}, {Donaldson}, {Duan},
  {Duckworth}, {Eftekharzadeh}, {Eisenstein}, {Etourneau}, {Fagrelius},
  {Farihi}, {Fitzpatrick}, {Font-Ribera}, {Fulmer}, {G{\"a}nsicke},
  {Gaztanaga}, {George}, {Gerdes}, {Gontcho}, {Gorgoni}, {Green}, {Guy},
  {Harmer}, {Hernandez}, {Honscheid}, {Huang}, {James}, {Jannuzi}, {Jiang},
  {Joyce}, {Karcher}, {Karkar}, {Kehoe}, {Kneib}, {Kueter-Young}, {Lan},
  {Lauer}, {Le Guillou}, {Le Van Suu}, {Lee}, {Lesser}, {Perreault Levasseur},
  {Li}, {Mann}, {Marshall}, {Mart{\'\i}nez-V{\'a}zquez}, {Martini}, {du Mas des
  Bourboux}, {McManus}, {Meier}, {M{\'e}nard}, {Metcalfe},
  {Mu{\~n}oz-Guti{\'e}rrez}, {Najita}, {Napier}, {Narayan}, {Newman}, {Nie},
  {Nord}, {Norman}, {Olsen}, {Paat}, {Palanque-Delabrouille}, {Peng},
  {Poppett}, {Poremba}, {Prakash}, {Rabinowitz}, {Raichoor}, {Rezaie},
  {Robertson}, {Roe}, {Ross}, {Ross}, {Rudnick}, {Safonova}, {Saha},
  {S{\'a}nchez}, {Savary}, {Schweiker}, {Scott}, {Seo}, {Shan}, {Silva},
  {Slepian}, {Soto}, {Sprayberry}, {Staten}, {Stillman}, {Stupak}, {Summers},
  {Sien Tie}, {Tirado}, {Vargas-Maga{\~n}a}, {Vivas}, {Wechsler}, {Williams},
  {Yang}, {Yang}, {Yapici}, {Zaritsky}, {Zenteno}, {Zhang}, {Zhang}, {Zhou}, \&
  {Zhou}}]{legacy}
{Dey}, A., {Schlegel}, D.~J., {Lang}, D., {et~al.} 2019, \aj, 157, 168,
  \dodoi{10.3847/1538-3881/ab089d}

\bibitem[{{Dye} {et~al.}(2018){Dye}, {Lawrence}, {Read}, {Fan}, {Kerr},
  {Varricatt}, {Furnell}, {Edge}, {Irwin}, {Hambly}, {Lucas}, {Almaini},
  {Chambers}, {Green}, {Hewett}, {Liu}, {McGreer}, {Best}, {Zhang}, {Sutorius},
  {Froebrich}, {Magnier}, {Hasinger}, {Lederer}, {Bold}, \& {Tedds}}]{uhs}
{Dye}, S., {Lawrence}, A., {Read}, M.~A., {et~al.} 2018, \mnras, 473, 5113,
  \dodoi{10.1093/mnras/stx2622}

\bibitem[{{Faber} \& {Jackson}(1976)}]{fj}
{Faber}, S.~M., \& {Jackson}, R.~E. 1976, \apj, 204, 668,
  \dodoi{10.1086/154215}

\bibitem[{{Fan} {et~al.}(2019){Fan}, {Wang}, {Yang}, {Keeton}, {Yue},
  {Zabludoff}, {Bian}, {Bonaglia}, {Georgiev}, {Hennawi}, {Li}, {McGreer},
  {Naidu}, {Pacucci}, {Rabien}, {Thompson}, {Venemans}, {Walter}, {Wang}, \&
  {Wu}}]{fan19}
{Fan}, X., {Wang}, F., {Yang}, J., {et~al.} 2019, \apjl, 870, L11,
  \dodoi{10.3847/2041-8213/aaeffe}

\bibitem[{{Focardi} \& {Malavasi}(2012)}]{forcardi12}
{Focardi}, P., \& {Malavasi}, N. 2012, \apj, 756, 117,
  \dodoi{10.1088/0004-637X/756/2/117}

\bibitem[{{Geng} {et~al.}(2021){Geng}, {Cao}, {Liu}, {Liu}, {Biesiada}, \&
  {Lian}}]{geng21}
{Geng}, S., {Cao}, S., {Liu}, Y., {et~al.} 2021, \mnras, 503, 1319,
  \dodoi{10.1093/mnras/stab519}

\bibitem[{{Hasan} \& {Crocker}(2019)}]{hasan19}
{Hasan}, F., \& {Crocker}, A. 2019, arXiv e-prints, arXiv:1904.00486.
\newblock \doarXiv{1904.00486}

\bibitem[{{Hilbert} {et~al.}(2008){Hilbert}, {White}, {Hartlap}, \&
  {Schneider}}]{hilbert08}
{Hilbert}, S., {White}, S. D.~M., {Hartlap}, J., \& {Schneider}, P. 2008,
  \mnras, 386, 1845, \dodoi{10.1111/j.1365-2966.2008.13190.x}

\bibitem[{{Ivezi{\'c}} {et~al.}(2019){Ivezi{\'c}}, {Kahn}, {Tyson}, {Abel},
  {Acosta}, {Allsman}, {Alonso}, {AlSayyad}, {Anderson}, {Andrew}, {Angel},
  {Angeli}, {Ansari}, {Antilogus}, {Araujo}, {Armstrong}, {Arndt}, {Astier},
  {Aubourg}, {Auza}, {Axelrod}, {Bard}, {Barr}, {Barrau}, {Bartlett}, {Bauer},
  {Bauman}, {Baumont}, {Bechtol}, {Bechtol}, {Becker}, {Becla}, {Beldica},
  {Bellavia}, {Bianco}, {Biswas}, {Blanc}, {Blazek}, {Blandford}, {Bloom},
  {Bogart}, {Bond}, {Booth}, {Borgland}, {Borne}, {Bosch}, {Boutigny},
  {Brackett}, {Bradshaw}, {Brandt}, {Brown}, {Bullock}, {Burchat}, {Burke},
  {Cagnoli}, {Calabrese}, {Callahan}, {Callen}, {Carlin}, {Carlson},
  {Chandrasekharan}, {Charles-Emerson}, {Chesley}, {Cheu}, {Chiang}, {Chiang},
  {Chirino}, {Chow}, {Ciardi}, {Claver}, {Cohen-Tanugi}, {Cockrum}, {Coles},
  {Connolly}, {Cook}, {Cooray}, {Covey}, {Cribbs}, {Cui}, {Cutri}, {Daly},
  {Daniel}, {Daruich}, {Daubard}, {Daues}, {Dawson}, {Delgado}, {Dellapenna},
  {de Peyster}, {de Val-Borro}, {Digel}, {Doherty}, {Dubois},
  {Dubois-Felsmann}, {Durech}, {Economou}, {Eifler}, {Eracleous}, {Emmons},
  {Fausti Neto}, {Ferguson}, {Figueroa}, {Fisher-Levine}, {Focke}, {Foss},
  {Frank}, {Freemon}, {Gangler}, {Gawiser}, {Geary}, {Gee}, {Geha}, {Gessner},
  {Gibson}, {Gilmore}, {Glanzman}, {Glick}, {Goldina}, {Goldstein}, {Goodenow},
  {Graham}, {Gressler}, {Gris}, {Guy}, {Guyonnet}, {Haller}, {Harris},
  {Hascall}, {Haupt}, {Hernandez}, {Herrmann}, {Hileman}, {Hoblitt}, {Hodgson},
  {Hogan}, {Howard}, {Huang}, {Huffer}, {Ingraham}, {Innes}, {Jacoby}, {Jain},
  {Jammes}, {Jee}, {Jenness}, {Jernigan}, {Jevremovi{\'c}}, {Johns}, {Johnson},
  {Johnson}, {Jones}, {Juramy-Gilles}, {Juri{\'c}}, {Kalirai}, {Kallivayalil},
  {Kalmbach}, {Kantor}, {Karst}, {Kasliwal}, {Kelly}, {Kessler}, {Kinnison},
  {Kirkby}, {Knox}, {Kotov}, {Krabbendam}, {Krughoff}, {Kub{\'a}nek},
  {Kuczewski}, {Kulkarni}, {Ku}, {Kurita}, {Lage}, {Lambert}, {Lange},
  {Langton}, {Le Guillou}, {Levine}, {Liang}, {Lim}, {Lintott}, {Long},
  {Lopez}, {Lotz}, {Lupton}, {Lust}, {MacArthur}, {Mahabal}, {Mandelbaum},
  {Markiewicz}, {Marsh}, {Marshall}, {Marshall}, {May}, {McKercher}, {McQueen},
  {Meyers}, {Migliore}, {Miller}, {Mills}, {Miraval}, {Moeyens}, {Moolekamp},
  {Monet}, {Moniez}, {Monkewitz}, {Montgomery}, {Morrison}, {Mueller},
  {Muller}, {Mu{\~n}oz Arancibia}, {Neill}, {Newbry}, {Nief}, {Nomerotski},
  {Nordby}, {O'Connor}, {Oliver}, {Olivier}, {Olsen}, {O'Mullane}, {Ortiz},
  {Osier}, {Owen}, {Pain}, {Palecek}, {Parejko}, {Parsons}, {Pease},
  {Peterson}, {Peterson}, {Petravick}, {Libby Petrick}, {Petry},
  {Pierfederici}, {Pietrowicz}, {Pike}, {Pinto}, {Plante}, {Plate}, {Plutchak},
  {Price}, {Prouza}, {Radeka}, {Rajagopal}, {Rasmussen}, {Regnault}, {Reil},
  {Reiss}, {Reuter}, {Ridgway}, {Riot}, {Ritz}, {Robinson}, {Roby}, {Roodman},
  {Rosing}, {Roucelle}, {Rumore}, {Russo}, {Saha}, {Sassolas}, {Schalk},
  {Schellart}, {Schindler}, {Schmidt}, {Schneider}, {Schneider}, {Schoening},
  {Schumacher}, {Schwamb}, {Sebag}, {Selvy}, {Sembroski}, {Seppala}, {Serio},
  {Serrano}, {Shaw}, {Shipsey}, {Sick}, {Silvestri}, {Slater}, {Smith},
  {Smith}, {Sobhani}, {Soldahl}, {Storrie-Lombardi}, {Stover}, {Strauss},
  {Street}, {Stubbs}, {Sullivan}, {Sweeney}, {Swinbank}, {Szalay}, {Takacs},
  {Tether}, {Thaler}, {Thayer}, {Thomas}, {Thornton}, {Thukral}, {Tice},
  {Trilling}, {Turri}, {Van Berg}, {Vanden Berk}, {Vetter}, {Virieux},
  {Vucina}, {Wahl}, {Walkowicz}, {Walsh}, {Walter}, {Wang}, {Wang}, {Warner},
  {Wiecha}, {Willman}, {Winters}, {Wittman}, {Wolff}, {Wood-Vasey}, {Wu},
  {Xin}, {Yoachim}, \& {Zhan}}]{lsst}
{Ivezi{\'c}}, {\v{Z}}., {Kahn}, S.~M., {Tyson}, J.~A., {et~al.} 2019, \apj,
  873, 111, \dodoi{10.3847/1538-4357/ab042c}

\bibitem[{{Jiang} {et~al.}(2016){Jiang}, {McGreer}, {Fan}, {Strauss},
  {Ba{\~n}ados}, {Becker}, {Bian}, {Farnsworth}, {Shen}, {Wang}, {Wang},
  {Wang}, {White}, {Wu}, {Wu}, {Yang}, \& {Yang}}]{jiang16}
{Jiang}, L., {McGreer}, I.~D., {Fan}, X., {et~al.} 2016, \apj, 833, 222,
  \dodoi{10.3847/1538-4357/833/2/222}

\bibitem[{{Kormann} {et~al.}(1994){Kormann}, {Schneider}, \&
  {Bartelmann}}]{sie}
{Kormann}, R., {Schneider}, P., \& {Bartelmann}, M. 1994, \aap, 284, 285

\bibitem[{{Korytov} {et~al.}(2019){Korytov}, {Hearin}, {Kovacs}, {Larsen},
  {Rangel}, {Hollowed}, {Benson}, {Heitmann}, {Mao}, {Bahmanyar}, {Chang},
  {Campbell}, {DeRose}, {Finkel}, {Frontiere}, {Gawiser}, {Habib}, {Joachimi},
  {Lanusse}, {Li}, {Mandelbaum}, {Morrison}, {Newman}, {Pope}, {Rykoff},
  {Simet}, {To}, {Vikraman}, {Wechsler}, {White}, \& {(The LSST Dark Energy
  Science Collaboration}}]{dc2}
{Korytov}, D., {Hearin}, A., {Kovacs}, E., {et~al.} 2019, \apjs, 245, 26,
  \dodoi{10.3847/1538-4365/ab510c}

\bibitem[{{Kulkarni} {et~al.}(2019){Kulkarni}, {Worseck}, \&
  {Hennawi}}]{kulkarni19}
{Kulkarni}, G., {Worseck}, G., \& {Hennawi}, J.~F. 2019, \mnras, 488, 1035,
  \dodoi{10.1093/mnras/stz1493}

\bibitem[{{Mason} {et~al.}(2015){Mason}, {Treu}, {Schmidt}, {Collett},
  {Trenti}, {Marshall}, {Barone-Nugent}, {Bradley}, {Stiavelli}, \&
  {Wyithe}}]{mason15}
{Mason}, C.~A., {Treu}, T., {Schmidt}, K.~B., {et~al.} 2015, \apj, 805, 79,
  \dodoi{10.1088/0004-637X/805/1/79}

\bibitem[{{Matsuoka} {et~al.}(2018){Matsuoka}, {Strauss}, {Kashikawa}, {Onoue},
  {Iwasawa}, {Tang}, {Lee}, {Imanishi}, {Nagao}, {Akiyama}, {Asami}, {Bosch},
  {Furusawa}, {Goto}, {Gunn}, {Harikane}, {Ikeda}, {Izumi}, {Kawaguchi},
  {Kato}, {Kikuta}, {Kohno}, {Komiyama}, {Lupton}, {Minezaki}, {Miyazaki},
  {Murayama}, {Niida}, {Nishizawa}, {Noboriguchi}, {Oguri}, {Ono}, {Ouchi},
  {Price}, {Sameshima}, {Schulze}, {Shirakata}, {Silverman}, {Sugiyama},
  {Tait}, {Takada}, {Takata}, {Tanaka}, {Toba}, {Utsumi}, {Wang}, \&
  {Yamashita}}]{matsuoka18}
{Matsuoka}, Y., {Strauss}, M.~A., {Kashikawa}, N., {et~al.} 2018, \apj, 869,
  150, \dodoi{10.3847/1538-4357/aaee7a}

\bibitem[{{McGreer} {et~al.}(2018){McGreer}, {Fan}, {Jiang}, \&
  {Cai}}]{mcgreer18}
{McGreer}, I.~D., {Fan}, X., {Jiang}, L., \& {Cai}, Z. 2018, \aj, 155, 131,
  \dodoi{10.3847/1538-3881/aaaab4}

\bibitem[{{McGreer} {et~al.}(2010){McGreer}, {Hall}, {Fan}, {Bian}, {Inada},
  {Oguri}, {Strauss}, {Schneider}, \& {Farnsworth}}]{mcgreer10}
{McGreer}, I.~D., {Hall}, P.~B., {Fan}, X., {et~al.} 2010, \aj, 140, 370,
  \dodoi{10.1088/0004-6256/140/2/370}

\bibitem[{{McGreer} {et~al.}(2013){McGreer}, {Jiang}, {Fan}, {Richards},
  {Strauss}, {Ross}, {White}, {Shen}, {Schneider}, {Myers}, {Brandt}, {DeGraf},
  {Glikman}, {Ge}, \& {Streblyanska}}]{mcgreer13}
{McGreer}, I.~D., {Jiang}, L., {Fan}, X., {et~al.} 2013, \apj, 768, 105,
  \dodoi{10.1088/0004-637X/768/2/105}

\bibitem[{{McMahon} {et~al.}(2013){McMahon}, {Banerji}, {Gonzalez}, {Koposov},
  {Bejar}, {Lodieu}, {Rebolo}, \& {VHS Collaboration}}]{vhs}
{McMahon}, R.~G., {Banerji}, M., {Gonzalez}, E., {et~al.} 2013, The Messenger,
  154, 35

\bibitem[{{More} {et~al.}(2016){More}, {Oguri}, {Kayo}, {Zinn}, {Strauss},
  {Santiago}, {Mosquera}, {Inada}, {Kochanek}, {Rusu}, {Brownstein}, {da
  Costa}, {Kneib}, {Maia}, {Quimby}, {Schneider}, {Streblyanska}, \&
  {York}}]{more16}
{More}, A., {Oguri}, M., {Kayo}, I., {et~al.} 2016, \mnras, 456, 1595,
  \dodoi{10.1093/mnras/stv2813}

\bibitem[{{Oguri}(2010)}]{glafic}
{Oguri}, M. 2010, {glafic: Software Package for Analyzing Gravitational
  Lensing}.
\newblock \doeprint{1010.012}

\bibitem[{{Oguri} \& {Marshall}(2010)}]{om10}
{Oguri}, M., \& {Marshall}, P.~J. 2010, \mnras, 405, 2579,
  \dodoi{10.1111/j.1365-2966.2010.16639.x}

\bibitem[{{Pacucci} \& {Loeb}(2019)}]{pacucci19}
{Pacucci}, F., \& {Loeb}, A. 2019, \apjl, 870, L12,
  \dodoi{10.3847/2041-8213/aaf86a}

\bibitem[{{Pei}(1995)}]{pei95}
{Pei}, Y.~C. 1995, \apj, 438, 623, \dodoi{10.1086/175105}

\bibitem[{{Richards} {et~al.}(2002){Richards}, {Fan}, {Newberg}, {Strauss},
  {Vanden Berk}, {Schneider}, {Yanny}, {Boucher}, {Burles}, {Frieman}, {Gunn},
  {Hall}, {Ivezi{\'c}}, {Kent}, {Loveday}, {Lupton}, {Rockosi}, {Schlegel},
  {Stoughton}, {SubbaRao}, \& {York}}]{richards02}
{Richards}, G.~T., {Fan}, X., {Newberg}, H.~J., {et~al.} 2002, \aj, 123, 2945,
  \dodoi{10.1086/340187}

\bibitem[{{Richards} {et~al.}(2004){Richards}, {Strauss}, {Pindor}, {Haiman},
  {Fan}, {Eisenstein}, {Schneider}, {Bahcall}, {Brinkmann}, \&
  {Brunner}}]{richards2004}
{Richards}, G.~T., {Strauss}, M.~A., {Pindor}, B., {et~al.} 2004, \aj, 127,
  1305, \dodoi{10.1086/381906}

\bibitem[{{Richards} {et~al.}(2006){Richards}, {Haiman}, {Pindor}, {Strauss},
  {Fan}, {Eisenstein}, {Schneider}, {Bahcall}, {Brinkmann}, \&
  {Fukugita}}]{richards06}
{Richards}, G.~T., {Haiman}, Z., {Pindor}, B., {et~al.} 2006, \aj, 131, 49,
  \dodoi{10.1086/498063}

\bibitem[{{Scaramella} {et~al.}(2021){Scaramella}, {Amiaux}, {Mellier},
  {Burigana}, {Carvalho}, {Cuillandre}, {Da Silva}, {Derosa}, {Dinis},
  {Maiorano}, {Maris}, {Tereno}, {Laureijs}, {Boenke}, {Buenadicha}, {Dupac},
  {Gaspar Venancio}, {G{\'o}mez-{\'A}lvarez}, {Hoar}, {Alvarez}, {Racca},
  {Saavedra-Criado}, {Schwartz}, {Vavrek}, {Schirmer}, {Aussel}, {Azzollini},
  {Cardone}, {Cropper}, {Ealet}, {Garilli}, {Gillard}, {Granett}, {Guzzo},
  {Hoekstra}, {Jahnke}, {Kitching}, {Meneghetti}, {Miller}, {Nakajima},
  {Niemi}, {Pasian}, {Percival}, {Sauvage}, {Scodeggio}, {Wachter}, {Zacchei},
  {Aghanim}, {Amara}, {Auphan}, {Auricchio}, {Awan}, {Balestra}, {Bender},
  {Bodendorf}, {Bonino}, {Branchini}, {Brau-Nogue}, {Brescia}, {Candini},
  {Capobianco}, {Carbone}, {Carlberg}, {Carretero}, {Casas}, {Castander},
  {Castellano}, {Cavuoti}, {Cimatti}, {Cledassou}, {Congedo}, {Conselice},
  {Conversi}, {Copin}, {Corcione}, {Costille}, {Courbin}, {Degaudenzi},
  {Douspis}, {Dubath}, {Duncan}, {Dusini}, {Farrens}, {Ferriol}, {Fosalba},
  {Fourmanoit}, {Frailis}, {Franceschi}, {Franzetti}, {Fumana}, {Gillis},
  {Giocoli}, {Grazian}, {Grupp}, {Haugan}, {Holmes}, {Hormuth}, {Hudelot},
  {Kermiche}, {Kiessling}, {Kilbinger}, {Kohley}, {Kubik}, {K{\"u}mmel},
  {Kunz}, {Kurki-Suonio}, {Ligori}, {Lilje}, {Lloro}, {Mansutti}, {Marggraf},
  {Markovic}, {Marulli}, {Massey}, {Maurogordato}, {Melchior}, {Merlin},
  {Meylan}, {Mohr}, {Moresco}, {Morin}, {Moscardini}, {Munari}, {Nichol},
  {Padilla}, {Paltani}, {Peacock}, {Pedersen}, {Pettorino}, {Pires}, {Poncet},
  {Popa}, {Pozzetti}, {Raison}, {Rebolo}, {Rhodes}, {Rix}, {Roncarelli},
  {Rossetti}, {Saglia}, {Schneider}, {Schrabback}, {Secroun}, {Seidel},
  {Serrano}, {Sirignano}, {Sirri}, {Skottfelt}, {Stanco}, {Starck},
  {Tallada-Cresp{\'\i}}, {Tavagnacco}, {Taylor}, {Teplitz}, {Toledo-Moreo},
  {Torradeflot}, {Trifoglio}, {Valentijn}, {Valenziano}, {Verdoes Kleijn},
  {Wang}, {Welikala}, {Weller}, {Wetzstein}, {Zamorani}, {Zoubian}, {Andreon},
  {Baldi}, {Bardelli}, {Boucaud}, {Camera}, {Fabbian}, {Farinelli},
  {Graci{\'a}-Carpio}, {Maino}, {Medinaceli}, {Mei}, {Neissner}, {Polenta},
  {Renzi}, {Romelli}, {Rosset}, {Sureau}, {Tenti}, {Vassallo}, {Zucca},
  {Baccigalupi}, {Balaguera-Antol{\'\i}nez}, {Battaglia}, {Biviano}, {Borgani},
  {Bozzo}, {Cabanac}, {Cappi}, {Casas}, {Castignani}, {Colodro-Conde},
  {Coupon}, {Courtois}, {Cuby}, {de la Torre}, {Desai}, {Di Ferdinando},
  {Dole}, {Fabricius}, {Farina}, {Ferreira}, {Finelli}, {Flose-Reimberg},
  {Fotopoulou}, {Galeotta}, {Ganga}, {Gozaliasl}, {Hook}, {Keihanen},
  {Kirkpatrick}, {Liebing}, {Lindholm}, {Mainetti}, {Martinelli}, {Martinet},
  {Maturi}, {McCracken}, {Metcalf}, {Morgante}, {Nightingale}, {Nucita},
  {Patrizii}, {Potter}, {Riccio}, {S{\'a}nchez}, {Sapone}, {Schewtschenko},
  {Schultheis}, {Scottez}, {Teyssier}, {Tutusaus}, {Valiviita}, {Viel},
  {Vriend}, \& {Whittaker}}]{euclid}
{Scaramella}, R., {Amiaux}, J., {Mellier}, Y., {et~al.} 2021, arXiv e-prints,
  arXiv:2108.01201.
\newblock \doarXiv{2108.01201}

\bibitem[{{Schneider} {et~al.}(1992){Schneider}, {Ehlers}, \&
  {Falco}}]{gravlens}
{Schneider}, P., {Ehlers}, J., \& {Falco}, E.~E. 1992, {Gravitational Lenses},
  \dodoi{10.1007/978-3-662-03758-4}

\bibitem[{{Sohn} {et~al.}(2017){Sohn}, {Zahid}, \& {Geller}}]{sohn17}
{Sohn}, J., {Zahid}, H.~J., \& {Geller}, M.~J. 2017, \apj, 845, 73,
  \dodoi{10.3847/1538-4357/aa7de3}

\bibitem[{{Springel} {et~al.}(2005){Springel}, {White}, {Jenkins}, {Frenk},
  {Yoshida}, {Gao}, {Navarro}, {Thacker}, {Croton}, {Helly}, {Peacock}, {Cole},
  {Thomas}, {Couchman}, {Evrard}, {Colberg}, \& {Pearce}}]{springel05}
{Springel}, V., {White}, S. D.~M., {Jenkins}, A., {et~al.} 2005, \nat, 435,
  629, \dodoi{10.1038/nature03597}

\bibitem[{{Wang} {et~al.}(2019){Wang}, {Yang}, {Fan}, {Wu}, {Yue}, {Li},
  {Bian}, {Jiang}, {Ba{\~n}ados}, {Schindler}, {Findlay}, {Davies}, {Decarli},
  {Farina}, {Green}, {Hennawi}, {Huang}, {Mazzuccheli}, {McGreer}, {Venemans},
  {Walter}, {Dye}, {Lyke}, {Myers}, \& {Haze Nunez}}]{wang19}
{Wang}, F., {Yang}, J., {Fan}, X., {et~al.} 2019, \apj, 884, 30,
  \dodoi{10.3847/1538-4357/ab2be5}

\bibitem[{{Wang} {et~al.}(2021){Wang}, {Yang}, {Fan}, {Hennawi}, {Barth},
  {Banados}, {Bian}, {Boutsia}, {Connor}, {Davies}, {Decarli}, {Eilers},
  {Farina}, {Green}, {Jiang}, {Li}, {Mazzucchelli}, {Nanni}, {Schindler},
  {Venemans}, {Walter}, {Wu}, \& {Yue}}]{wang21}
---. 2021, \apjl, 907, L1, \dodoi{10.3847/2041-8213/abd8c6}

\bibitem[{{Wyithe} \& {Loeb}(2002)}]{wyithe02}
{Wyithe}, J. S.~B., \& {Loeb}, A. 2002, \apj, 577, 57, \dodoi{10.1086/342181}

\bibitem[{{Wyithe} {et~al.}(2011){Wyithe}, {Yan}, {Windhorst}, \&
  {Mao}}]{wyithe11}
{Wyithe}, J. S.~B., {Yan}, H., {Windhorst}, R.~A., \& {Mao}, S. 2011, \nat,
  469, 181, \dodoi{10.1038/nature09619}

\bibitem[{{Yang} {et~al.}(2019{\natexlab{a}}){Yang}, {Wang}, {Fan}, {Wu},
  {Bian}, {Ba{\~n}ados}, {Yue}, {Schindler}, {Yang}, {Jiang}, {McGreer},
  {Green}, \& {Dye}}]{yang19}
{Yang}, J., {Wang}, F., {Fan}, X., {et~al.} 2019{\natexlab{a}}, \apj, 871, 199,
  \dodoi{10.3847/1538-4357/aaf858}

\bibitem[{{Yang} {et~al.}(2019{\natexlab{b}}){Yang}, {Venemans}, {Wang}, {Fan},
  {Novak}, {Decarli}, {Walter}, {Yue}, {Momjian}, {Keeton}, {Wang},
  {Zabludoff}, {Wu}, \& {Bian}}]{yang19b}
{Yang}, J., {Venemans}, B., {Wang}, F., {et~al.} 2019{\natexlab{b}}, \apj, 880,
  153, \dodoi{10.3847/1538-4357/ab2a02}

\bibitem[{{Yang} {et~al.}(2020){Yang}, {Wang}, {Fan}, {Hennawi}, {Davies},
  {Yue}, {Banados}, {Wu}, {Venemans}, {Barth}, {Bian}, {Boutsia}, {Decarli},
  {Farina}, {Green}, {Jiang}, {Li}, {Mazzucchelli}, \& {Walter}}]{yang20}
{Yang}, J., {Wang}, F., {Fan}, X., {et~al.} 2020, \apjl, 897, L14,
  \dodoi{10.3847/2041-8213/ab9c26}

\bibitem[{{York} {et~al.}(2000){York}, {Adelman}, {Anderson}, {Anderson},
  {Annis}, {Bahcall}, {Bakken}, {Barkhouser}, {Bastian}, {Berman}, {Boroski},
  {Bracker}, {Briegel}, {Briggs}, {Brinkmann}, {Brunner}, {Burles}, {Carey},
  {Carr}, {Castander}, {Chen}, {Colestock}, {Connolly}, {Crocker}, {Csabai},
  {Czarapata}, {Davis}, {Doi}, {Dombeck}, {Eisenstein}, {Ellman}, {Elms},
  {Evans}, {Fan}, {Federwitz}, {Fiscelli}, {Friedman}, {Frieman}, {Fukugita},
  {Gillespie}, {Gunn}, {Gurbani}, {de Haas}, {Haldeman}, {Harris}, {Hayes},
  {Heckman}, {Hennessy}, {Hindsley}, {Holm}, {Holmgren}, {Huang}, {Hull},
  {Husby}, {Ichikawa}, {Ichikawa}, {Ivezi{\'c}}, {Kent}, {Kim}, {Kinney},
  {Klaene}, {Kleinman}, {Kleinman}, {Knapp}, {Korienek}, {Kron}, {Kunszt},
  {Lamb}, {Lee}, {Leger}, {Limmongkol}, {Lindenmeyer}, {Long}, {Loomis},
  {Loveday}, {Lucinio}, {Lupton}, {MacKinnon}, {Mannery}, {Mantsch}, {Margon},
  {McGehee}, {McKay}, {Meiksin}, {Merelli}, {Monet}, {Munn}, {Narayanan},
  {Nash}, {Neilsen}, {Neswold}, {Newberg}, {Nichol}, {Nicinski}, {Nonino},
  {Okada}, {Okamura}, {Ostriker}, {Owen}, {Pauls}, {Peoples}, {Peterson},
  {Petravick}, {Pier}, {Pope}, {Pordes}, {Prosapio}, {Rechenmacher}, {Quinn},
  {Richards}, {Richmond}, {Rivetta}, {Rockosi}, {Ruthmansdorfer}, {Sandford},
  {Schlegel}, {Schneider}, {Sekiguchi}, {Sergey}, {Shimasaku}, {Siegmund},
  {Smee}, {Smith}, {Snedden}, {Stone}, {Stoughton}, {Strauss}, {Stubbs},
  {SubbaRao}, {Szalay}, {Szapudi}, {Szokoly}, {Thakar}, {Tremonti}, {Tucker},
  {Uomoto}, {Vanden Berk}, {Vogeley}, {Waddell}, {Wang}, {Watanabe},
  {Weinberg}, {Yanny}, {Yasuda}, \& {SDSS Collaboration}}]{sdss}
{York}, D.~G., {Adelman}, J., {Anderson}, John~E., J., {et~al.} 2000, \aj, 120,
  1579, \dodoi{10.1086/301513}

\bibitem[{{Yue} {et~al.}(2021){Yue}, {Yang}, {Fan}, {Wang}, {Spilker},
  {Georgiev}, {Keeton}, {Litke}, {Marrone}, {Walter}, {Wang}, {Wu}, {Venemans},
  \& {Zabludoff}}]{yue21}
{Yue}, M., {Yang}, J., {Fan}, X., {et~al.} 2021, \apj, 917, 99,
  \dodoi{10.3847/1538-4357/ac0af4}

\end{thebibliography}
\bibliographystyle{aasjournal}



\end{document}